\documentclass[pre,amsmath,amssymb,twocolumn,nofootinbib,floatfix]{revtex4-2}
\usepackage[colorlinks=true,linkcolor=blue,urlcolor=blue,citecolor=blue]{hyperref}
\usepackage{graphicx}
\usepackage{bm,color}
\usepackage{times}
\usepackage{color}
\usepackage{makecell}

\definecolor{orange}{rgb}{1,0.6, 0}
\definecolor{darkergreen}{rgb}{0,0.7,0}
\definecolor{grey}{rgb}{.4,.4,0.4}
\newcommand{\blue}{\color{blue}}

\renewcommand{\doi}[2]{\href{http://dx.doi.org/#1}{#2}}
\newcommand{\arxiv}[1]{\href{http://arxiv.org/abs/#1}{#1}}
\newcommand{\link}[2]{\href{http://#1}{#2}}

\newcommand{\Eq}[1]{Eq.~(\ref{#1})}
\newcommand{\Eqs}[1]{Eqs.~(\ref{#1})}
\newcommand{\eq}[1]{(\ref{#1})}
\newcommand{\half}{\frac12}
\newcommand{\bea}{\begin{eqnarray}}
\newcommand{\eea}{\end{eqnarray}}

\newcommand{\red}{\color{red}}

\newcommand{\rme}{\mathrm{e}}
\newcommand{\rmd}{\mathrm{d}}
\newcommand{\nn}{\nonumber}
\renewcommand{\epsilon}{\varepsilon}
\newcommand{\ca}[1]{{\cal #1}}
\newcommand{\be}{\begin{equation}}
\newcommand{\ee}{\end{equation}}

\usepackage[normalem]{ulem}

\graphicspath{{figures/},{}}

\begin{document}

\title{Force Correlations in Disordered Magnets}
\author{Cathelijne ter Burg$^1$, Felipe Bohn$^2$, Gianfranco Durin$^3$, Rubem Luis Sommer$^4$, Kay J\"org Wiese$^1$}
  \affiliation{\mbox{$^1$ Laboratoire de Physique de l'\'Ecole Normale Sup\'erieure, ENS, Universit\'e PSL, CNRS, Sorbonne Universit\'e,} \mbox{Universit\'e Paris-Diderot, Sorbonne Paris Cit\'e, 24 rue Lhomond, 75005 Paris, France}\\
  \mbox{$^2$  Departamento de F\'isica, Universidade Federal do Rio Grande do Norte, 59078-900 Natal, RN, Brazil}
  \mbox{$^3$ Istituto Nazionale di Ricerca Metrologica, strada delle Cacce 91, Torino, Italy}\\\mbox{$^4$ Centro Brasileiro de Pesquisas F\'isicas, Rua Dr. Xavier Sigaud 150, Urca, 22290-180 Rio de Janeiro, RJ, Brazil}}

\begin{abstract}
 We present  a {\em proof of principle} for the validity of the functional renormalization group, by measuring the   force correlations  in Barkhausen-noise experiments. Our samples are    soft ferromagnets  in two distinct universality classes, differing in the range of spin interactions, and the effects of eddy currents. We show that the force correlations have a universal form predicted by the functional renormalization group, distinct for  short-range and long-range elasticity, and mostly independent of eddy currents. In all cases correlations grow linearly at small distances, as in mean-field models, but in contrast to the latter are bounded at large distances. As a consequence,  avalanches   are anti-correlated. We derive bounds for these anti-correlations, which are saturated in the experiments, showing that the multiple domain walls in our samples  effectively behave as a single wall. 
\end{abstract} 

\maketitle

Each theory of disordered systems relies on specific assumptions, and    often their validity is checked only for  standard observables, such as the roughness exponent.  Measuring  its central ingredients would be a much more stringent test to discriminate between them. Two general theories have been   proposed: the Gaussian variational ansatz invoking replica-symmetry breaking~\cite{MezardParisiVirasoroBook,Parisi1979,Parisi1980b}, which is exact for fully connected models~\cite{Talagrand2011a,Talagrand2011b}, and the functional renormalization group (FRG) for short-ranged elastic systems~\cite{DSFisher1986,NattermannStepanowTangLeschhorn1992}, where the central ingredient is the effective force correlator. This correlator is the solution of a non-linear partial differential equation~\cite{DSFisher1986,NattermannStepanowTangLeschhorn1992,Wiese2021}, and can  experimentally be 
 extracted from the center-of-mass fluctuations of the interface.

To   prove the validity of the FRG  for disordered systems, we analyse the  domain-wall  motion in soft magnets (the Barkhausen noise)~\cite{BohnDurinCorreaMachadoDellaPaceChesmanSommer2018}, the oldest example of depinning and avalanche motion~\cite{Barkhausen1919,SethnaDahmenMyers2001,DurinZapperi2006b,Wiese2021}.
Standard   observables as  the avalanche size, duration~\cite{PerkovicDahmenSethna1995,DurinZapperi2000} and shape~\cite{ZapperiCastellanoColaioriDurin2005, PapanikolaouBohnSommerDurinZapperiSethna2011,LaursonIllaSantucciTallakstadyAlava2013,DurinBohnCorreaSommerDoussalWiese2016}  show  the existence of two universality classes differing in the kind and range of domain-wall interactions~\cite{DurinZapperi2000,DurinZapperi2006b}: amorphous materials with short-range (SR) interactions and polycrystals with long-range (LR) interactions, consequence of strong dipolar effects. In 3D magnets, the latter is   described by mean-field models pioneered in 1990 by Alessandro, Beatrice, Bertotti and Montorsi (ABBM)~\cite{AlessandroBeatriceBertottiMontorsi1990,AlessandroBeatriceBertottiMontorsi1990b,Colaiori2008}, where a  domain wall is represented by a single degree of freedom, its centre of mass, a.k.a.\ {\em mean field} (MF). 
For the SR class, key observables  as  the avalanche-size exponent $\tau\simeq 1.27$ differ from their MF prediction $\tau_{\rm MF}=3/2$, while they are accounted for by field-theoretic models~\cite{DSFisher1998,DobrinevskiLeDoussalWiese2014a,DobrinevskiPhD}.

In view of the solid evidence for exponents, a central question is whether experiments can contradict the ABBM model in a key prediction for LR magnets. 
We show that this is the case for the 
force correlator acting on the domain wall, or equivalently the correlator of its centre of mass. To understand this, consider the equation of motion of a $d$-dimensional interface with SR interactions,  
\begin{eqnarray}\label{eom}
\eta \partial_t u(x, t) &=& \nabla^2 u(x, t) + m^2 [w - u(x, t)] + F(x, u(x,t)),  \nn\\
w &=& vt. 
\end{eqnarray}
Here $w$ is proportional to the external applied field, increased very slowly, and $m^2$, usually denoted $k$, is the {\em demagnetization factor}~\cite{DurinZapperi2006b}.
Averaging \Eq{eom} over $x$, given $w$,  we get
$
\eta \dot u_w =  m^2 \left[w-u_w \right] +  F_w.$
Most of the time $\dot u_w=0$, and the position and force correlations are 
\be\label{obs}
 \hat \Delta_v(w-w') := \overline{\left[w-u_w \right] \left[w'-u_{w'} \right]}^{\rm c}   \simeq \frac1 {m^4}
\overline{F_w F_{w'}}^{\rm c} ,
\ee
{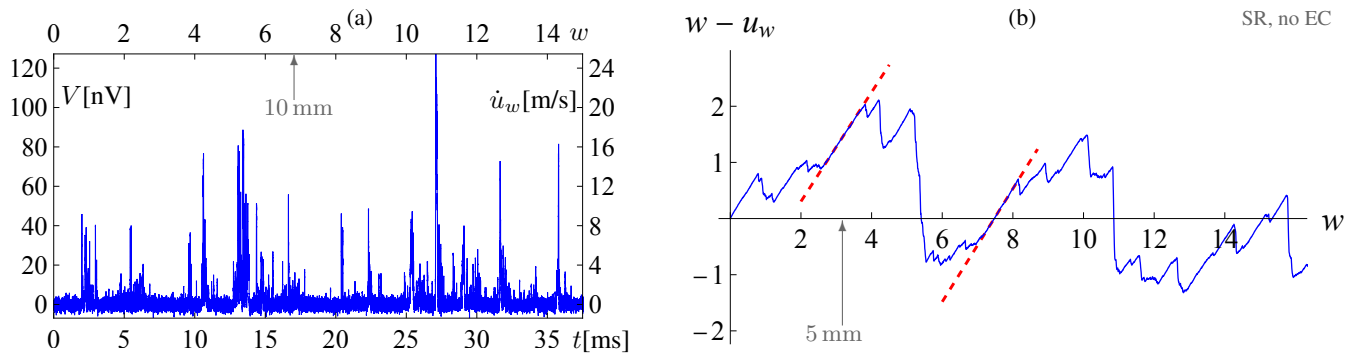
\begin{figure*}[t]
{\mbox{\color{black}\setlength{\unitlength}{1cm}\begin{picture}(17.7,4.55)
\put(-.1,-.1){{\includegraphics[width=.45\textwidth]{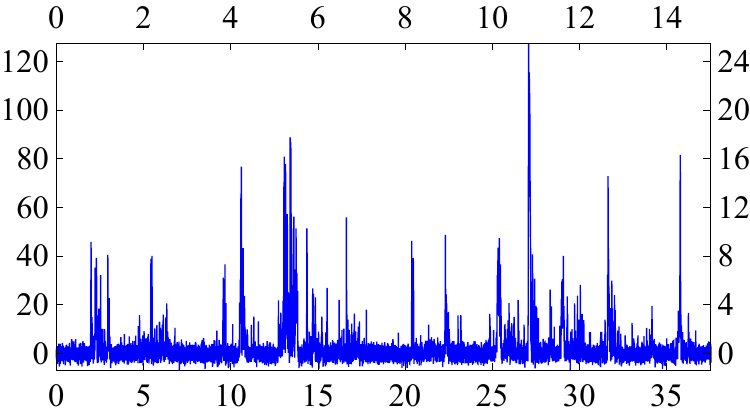}}}
\put(8.9,0){{\includegraphics[width=.49\textwidth]{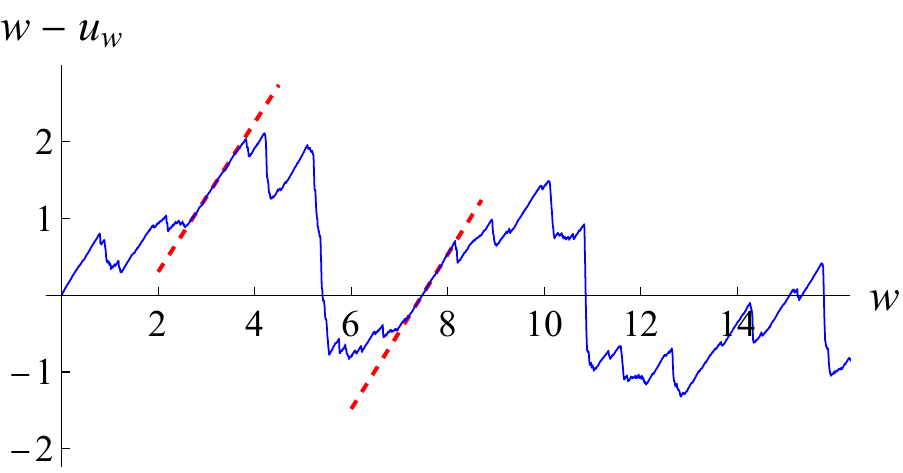}}}
\put(4.4,4.3){(a)}
\put(7.4,0){{\fontsize{10}{20}\selectfont $t$[ms]}}
\put(0.6,3.25){{\fontsize{10}{20}\selectfont $V$[nV]}}
\put(7.36,4.1){{\fontsize{10}{20}\selectfont  $w$}}
\put(6.3,3.2){{\fontsize{10}{20}\selectfont  $\dot{u}_w$[m/s]}} 
\put(13.2,4.3){(b)}
 \put(16.3,4.3){\footnotesize {\color{grey}SR, no EC}}
\put(3.3,3.1){\color{grey} $10\,\rm mm$}
\put(3.7,3.3){\color{grey}\vector(0,1){0.6}}
\put(10.5,0.1){\color{grey} $5\,\rm mm$}
\put(10.99,0.3){\color{grey}\vector(0,1){1.4}}
\end{picture}}}
\caption{Barkhausen noise in an amorphous FeSiB film (table \ref{tab1}).
 $\textbf{(a)}$ Voltage signal recorded in the experiment (left axis), and corresponding domain-wall velocity $\dot u_w$ (right axis), as a function of time (bottom axis) and $w$ (top axis).
$\textbf{(b)}$ 
The connected part of the interface position, $w-u_w$, obtained by  integrating $\dot u_w$.  
$w=1$  corresponds to   $ 2.5\,{\rm ms} \approx 1.5\,\rm mm$. Physical units are indicated by grey arrows.}
\label{FigSRNECForcedrops}
\end{figure*}}%
where  the overbar designates a disorder average and ${}^{\rm c}$ its connected part. In practice it is taken both over $w$ and runs.
 $\hat\Delta_v$ depends on the driving velocity.
Its zero-velocity limit  
\begin{align}
 \hat \Delta (w ) = \lim_{v \to 0} \hat \Delta_{v }(w ),  \label{DeltaDef}
\end{align}
 is   the central object of the FRG field theory~\cite{NattermannStepanowTangLeschhorn1992,NarayanDSFisher1993a,LeDoussal2006b,LeDoussalWiese2006a,Wiese2021}.
 
In an experiment, it is impossible to take $v\to 0$. 
The effect of {\blue $v>0$}  is to round the cusp $|\hat\Delta^\prime (0^+)|=\sigma $ (see \Eq{DeltaABBM}) in a boundary layer of size $\delta_{w} \sim v \tau$, where $\tau$ is the timescale set by the response function $R(t)\simeq \frac1\tau \rme^{-t/\tau}$ (see  Fig.~\ref{Fig2}(c) for an example).  
Ref.~\cite{terBurgWiese2020} shows that
\begin{align}
\hat \Delta_v(w) = \int_0^\infty {\rm d} t \int_0^\infty {\rm d} t^\prime R(t) R(t^\prime ) \hat\Delta(w - v(t - t^\prime))  \label{DeltaDefRounded-text}
\end{align}
can be  deconvoluted  to reconstruct $\hat\Delta(w)$ from the measured $\hat\Delta_v(w)$ (see App.~\ref{s:deconvolution}).  The result is  
\begin{align}
\hat\Delta(w)  = \hat \Delta_v(w) + \tau^2 \hat\Delta_{\dot{u}}(w) \label{OperatorUnfolding1-text},
\end{align}
where $\hat\Delta_{\dot u}(w)$ is the auto-correlation function of the   measured  $\dot u_w$. 
This allows us to extract {\blue $\hat \Delta(w)$} by plotting the r.h.s.\ and finding $\tau$ that best eliminates the rounding close to $w=0$. As shown below, Eq.\ \eqref{OperatorUnfolding1-text} allows us to remove a      boundary layer  of size $\delta_w = v \tau$, but it creates a smaller one of size $\delta^\prime_{w} = v \tau^\prime$,  see App.~\ref{s:Higher-order deconvolution}.

The ABBM model assumes that forces  $F_w$   perform a random walk, and as a consequence 
\be
\frac1{m^{4}}\frac{1}2 \overline{[F_w-F_{w'}]^2} = \hat  \Delta(0)- \hat  \Delta(w-w')\simeq \sigma |w-w'|. \label{DeltaABBM}
\ee
Field theory~\cite{ChauveLeDoussalWiese2000a,LeDoussalWieseChauve2002,Wiese2021} predicts  $\hat\Delta(0)-\hat\Delta(w)$ to grow linearly as \Eq{DeltaABBM} for small $w$, and to  saturate for large $w$, with distinct shapes in SR and LR systems (see App.~\ref{s:theory-Delta}). While this framework was tested in simulations~\cite{RossoLeDoussalWiese2006a,terBurgWiese2020}, and experiments on wetting~\cite{LeDoussalWieseMoulinetRolley2009} and RNA/DNA peeling~\cite{WieseBercyMelkonyanBizebard2019}, only with magnets we can consider two universality classes, and with  a large  statistics.

We analyze our experimental data   as follows. We start from    the Barkhausen-noise time series, proportional to the center-of-mass velocity $\dot u_w$ (See Fig.~\ref{FigSRNECForcedrops}(a)). The signal is characterized by bursts when the domain wall moves forward, and a vanishing signal when  it  is pinned i.e.\ at rest, combined with background  noise  (without noise $\dot u_w\ge 0$~\cite{Middleton1992}). This allows us to reconstruct the position of the center of mass $u_w$ (see \Eq{obs}), as depicted in~Fig.~\ref{FigSRNECForcedrops}(b). It is characterized by linearly increasing parts with slope 1, corresponding to an increasing magnetic field (i.e.\ $w$), followed by   drops in $w-u_w$ when the  wall moves forward.  
This allows us to reconstruct the  unknown  scale between $\dot u_w$ and the  voltage induced in the pickup  coil, reducing the scales in the experiment to a single one (see App.~\ref{s:Numerical methods}).

{\begin{figure*}[t]
{\setlength{\unitlength}{1cm}\begin{picture}(17.8,7.2)
\put(0,0){ \includegraphics[width=.49\textwidth]{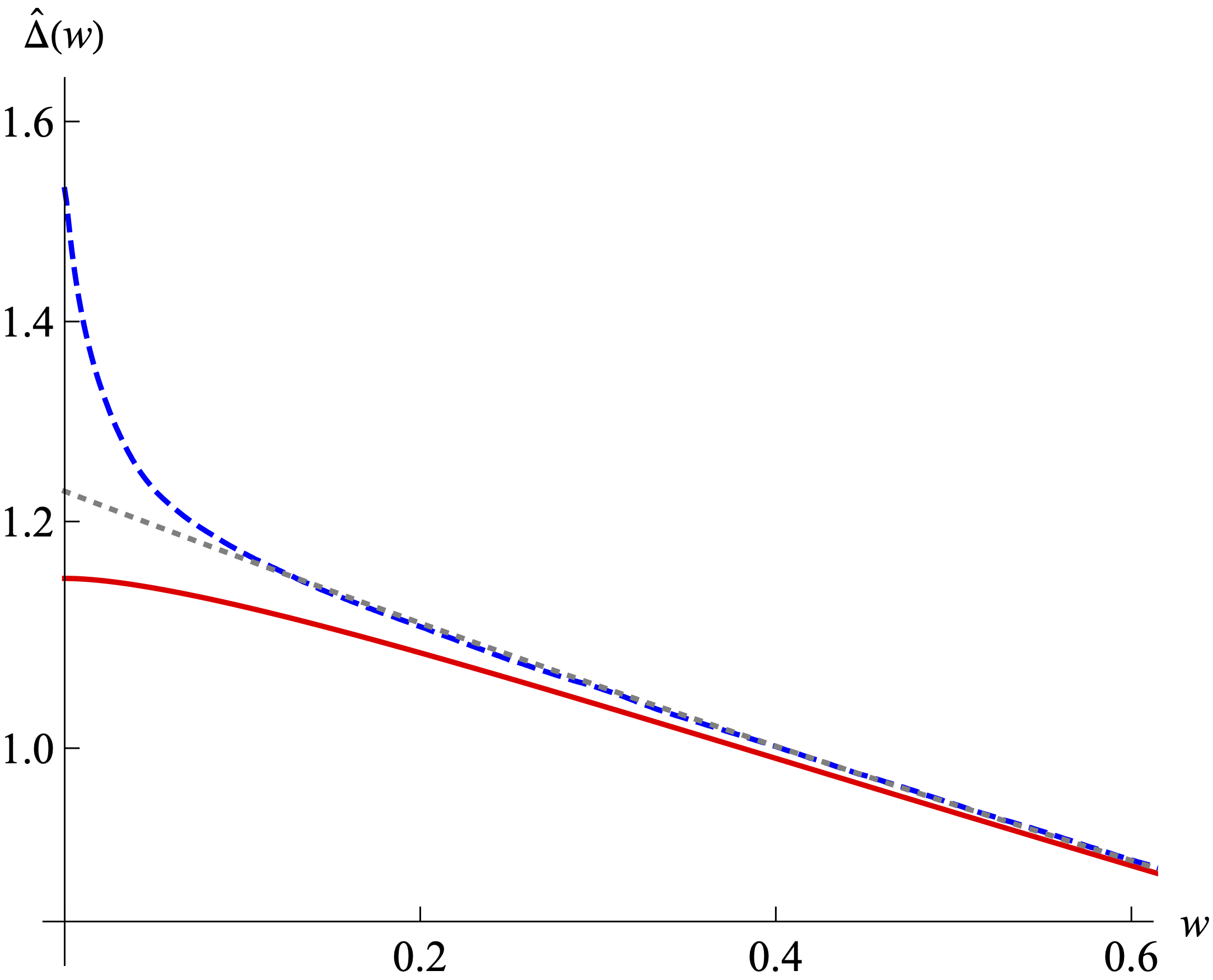}}
\put(8.9,0){ \includegraphics[width=.49\textwidth]{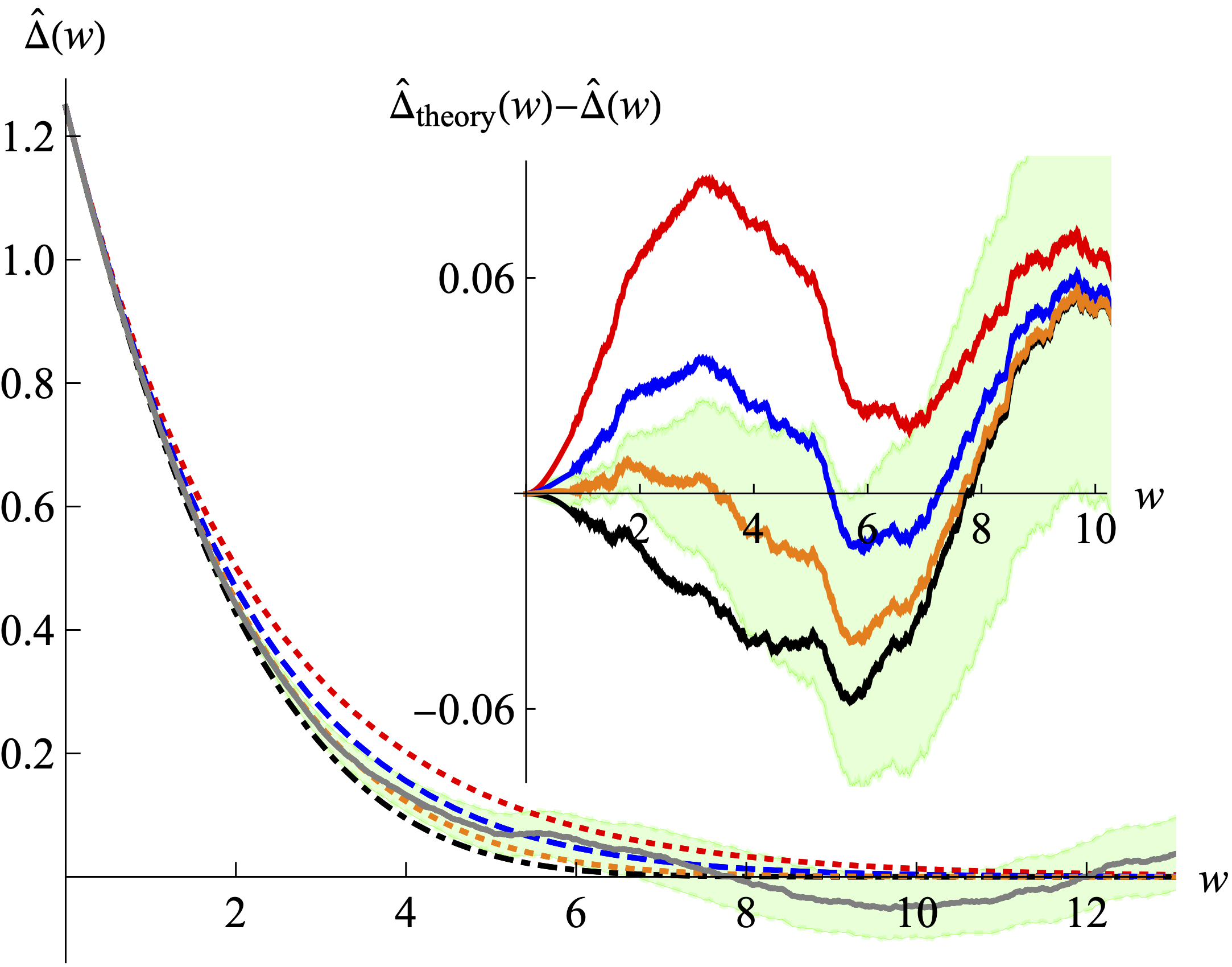}}
\put(4.4,7){(a)} \put(7,6.7){\footnotesize {\color{grey}SR, no EC}}
\put(12.2,7){(b)}
 \put(15.99,6.7){\footnotesize {\color{grey}SR, no EC}}
\put(1,1){\red raw data}
\put(1.3,1.3){\red\vector(0,5){1.47}}
\put(3.2,5.3){\color{blue} deconvoluted data}
\put(3.1,5.3){\color{blue}\vector(-3,-1){2.3}}
\put(3.1,4.){\color{grey} extrapolation}
\put(3,4.03){\color{grey}\vector(-3,-1){1.9}}
\put(11.,2.85){\color{red} exponential}
\put(11.4,2.7){\color{red}\vector(-0,-1){0.69}}
\put(13.1,1.5){\color{blue} $d=0$}
\put(13,1.5){\color{blue}\vector(-1,-1){0.45}}
\put(9.85,1.76){\color{black} 1-loop}
\put(9.9,1.4){\color{black} $d=d_{\rm c}$}
\put(10.75,1.5){\color{black}\vector(2,1){0.45}}
\put(10.17,1){\color{orange} 2-loop}
\put(10.2,0.73){\bfseries\color{orange} $ {\it d}=2$}
\put(11.07,0.9){\color{orange}\vector(2,1){0.70}}
\put(14.5,1.2){\color{grey} measurement}
\put(15.0,1.15){\color{grey}\vector(0,-1){0.7}}
\put(2.,-.3){\color{grey} $200\,\rm \mu m$}
\put(2.35,0){\color{grey}\vector(0,1){.4}}
\put(11.1,0){\color{grey} $5\,\rm mm$}
\put(11.4,0.28){\color{grey}\vector(0,1){.37}}
\end{picture}}
{\setlength{\unitlength}{1cm}\begin{picture}(17.8,7.4)
\put(0,0){ \includegraphics[width=.49\textwidth]{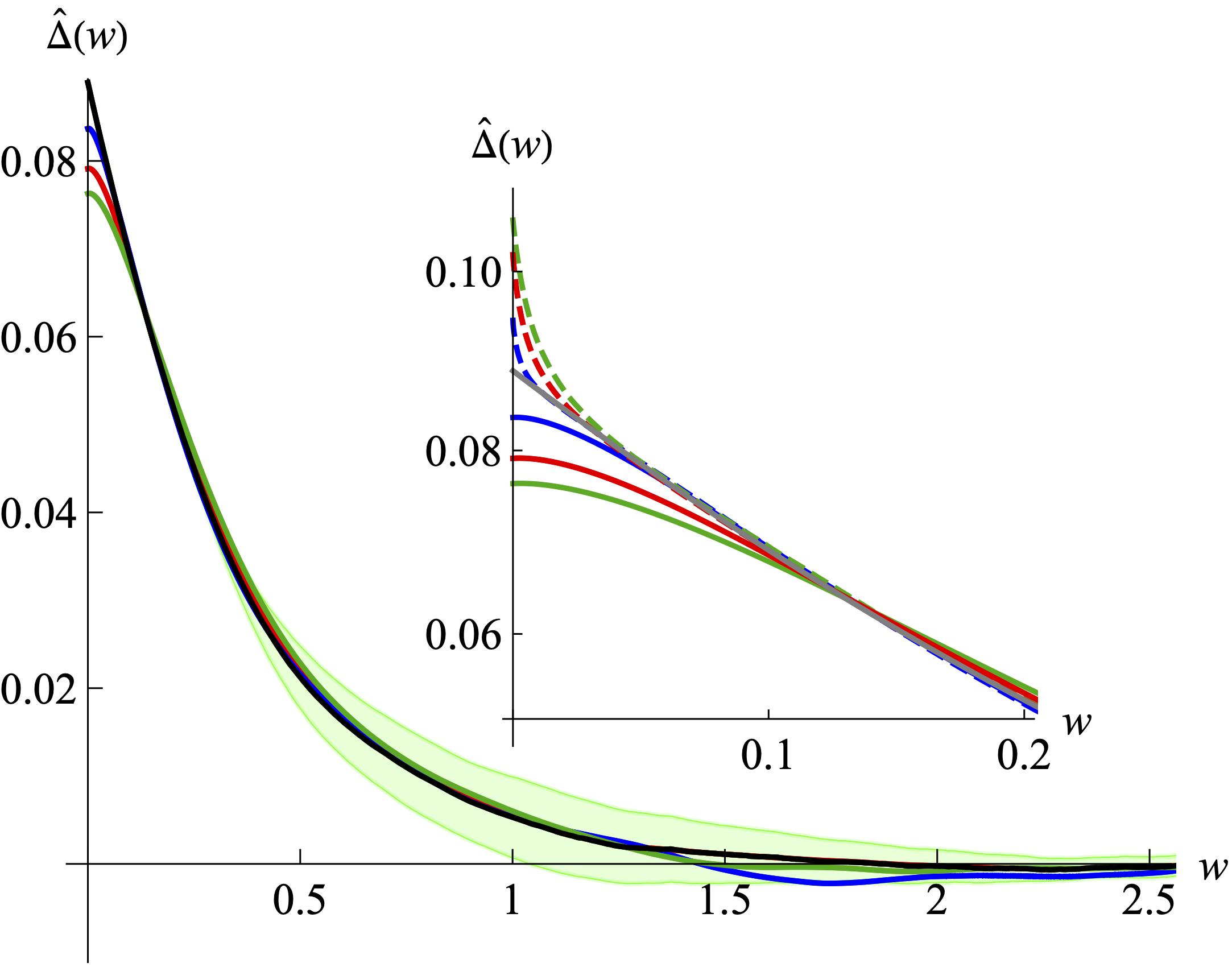}}
\put(8.9,0){ \includegraphics[width=.49\textwidth]{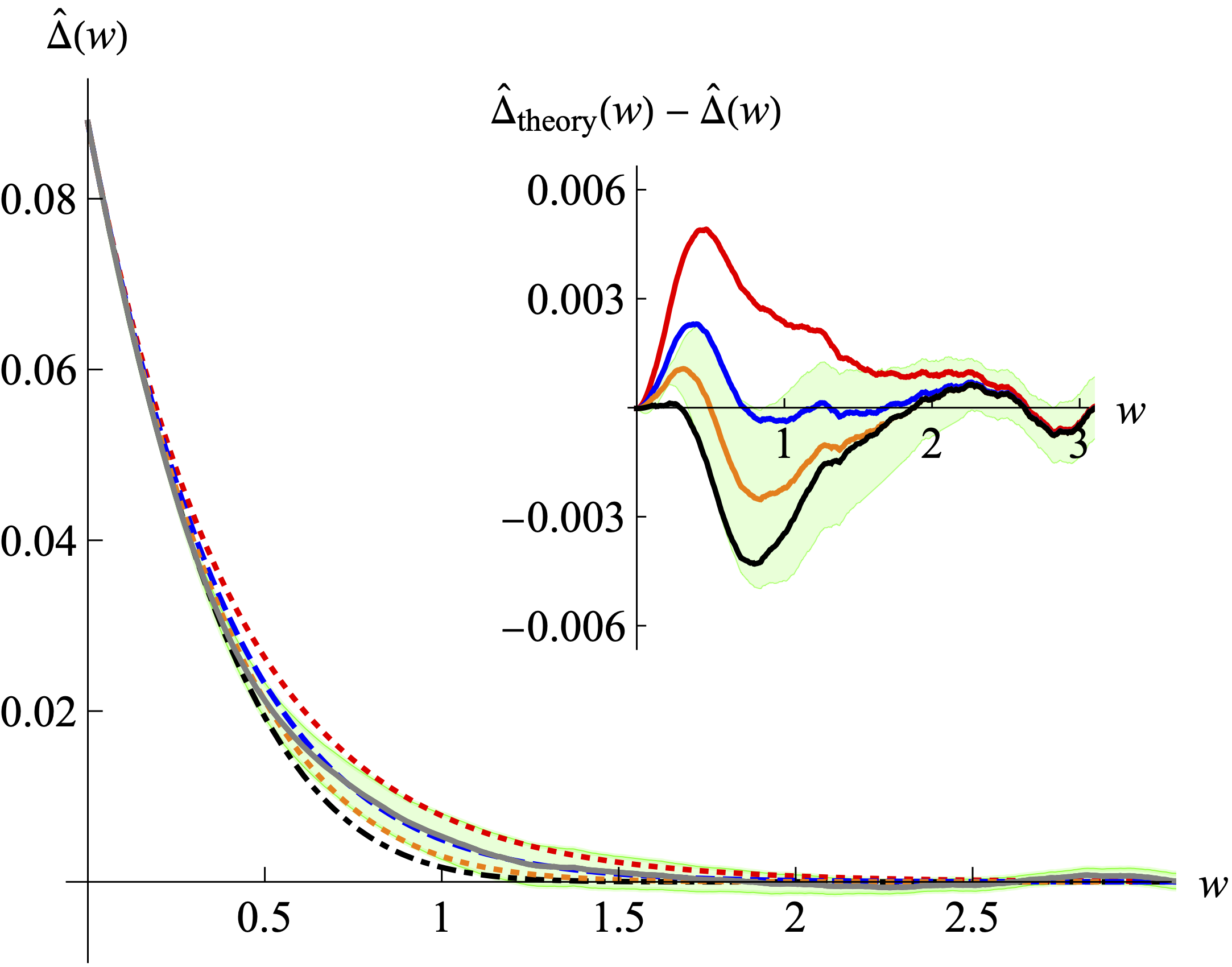}}
\put(4.4,7){(c)}
 \put(7.5,6.5){\footnotesize \color{grey}SR, EC}
 \put(16.5,6.5){\footnotesize \color{grey}SR, EC}
\put(13.2,7){(d)}
\put(4.8,5.93){\color{grey}extrapolation to $v=0$}
\put(6.23,5.87){\color{grey}\vector(-3,-2){2.45}}
\put(6.3,5.45){\color{blue} $v=1$}
\put(6.2,5.45){\color{blue}\vector(-3,-2){2.3}}
\put(6.3,4.8){\color{red} $v=2$}
\put(6.2,4.8){\color{red}\vector(-2,-1){2.3}}
\put(6.3,4.25){\color{darkergreen} $v=3$}
\put(6.2,4.25){\color{darkergreen}\vector(-2,-1){1.8}}
\put(11.,2.5){\color{red} exponential}
\put(11.4,2.4){\color{red}\vector(-0,-1){0.78}}
\put(11.,3.4){\color{blue} $d=0$}
\put(11.05,3.3){\color{blue}\vector(-1,-1){0.55}}
\put(9.7,1.7){\color{black} 1-loop}
\put(9.68,1.35){\color{black} $d=d_{\rm c}$}
\put(10.5,1.5){\color{black}\vector(2,1){0.4}}
\put(10.,0.95){\color{orange} 2-loop}
\put(10.03,0.65){\color{orange} $d=2$}
\put(10.9,0.85){\color{orange}\vector(2,1){0.6}}
\put(14.1,1.3){\color{grey} measurement}
\put(14.0,1.32){\color{grey}\vector(-4,-1){1.75}}
\put(2.8,0){\color{grey} $100\,\rm \mu m$}
\put(3,0.25){\color{grey}\vector(0,1){.5}}
\put(11,0){\color{grey} $100\,\rm \mu m$}
\put(11.5,0.2){\color{grey}\vector(0,1){.42}}
\end{picture}}
\caption{\textbf{(a)} Construction of 
$\hat\Delta(w)$ for   the FeSiB film   (SR,  no  ECs). In red the raw data. 
 In blue dashed, the result from Eq.~\eqref{OperatorUnfolding1-text}  using $\tau =0.17$. 
 In dotted gray the extrapolation to $w=0$.   
 \textbf{(b)}
Comparison of  $  \hat \Delta(w)$  using the   dotted gray  curve of  {(a)}, to   theory candidates, fixing  scales by $\hat\Delta(0)$ and $\hat\Delta'(0^+)$.  The 
latter are
from top to bottom: exponential  (red, dotted),   solution in $d = 0$~\cite{LeDoussalWiese2008a,terBurgWiese2020} (blue, dashed),   2-loop FRG   via Pad\'e  for $d = 2$  (orange, dotted),   1-loop FRG   
(black, dot-dashed).  Error bars  in green for 1-$\sigma$ confidence intervals. The inset shows theory minus data in the   same  color code, favoring   $d = 2$ FRG at two loops (with error bars for this curve only). 
{\bf (c)} Check of     deconvolution  \Eq{OperatorUnfolding1-text}, for  the FeCoB ribbon (SR, noticeable ECs),  at different driving velocities $v$, using the same time scale $\tau=0.025$; magnified in the inset. 
Apart from a small deviation for $v = 3$ they extrapolate to the same   function. $\textbf{(d)}$ Comparison of $\hat\Delta(w)$ from (c) to the   theory, using the   color code of Fig.~(b). 
The data is consistent with     2-loop FRG   in  $d = 2 $. 
$w=1$  corresponds to    $ 2.5\,{\rm ms} \approx 1.5\,\rm mm$   for (a)-(b), and to $ 0.2 \,{\rm s}\approx 135\,\mu\rm m$ for (c)-(d), see grey arrows.
}
\label{Fig2}
\end{figure*}
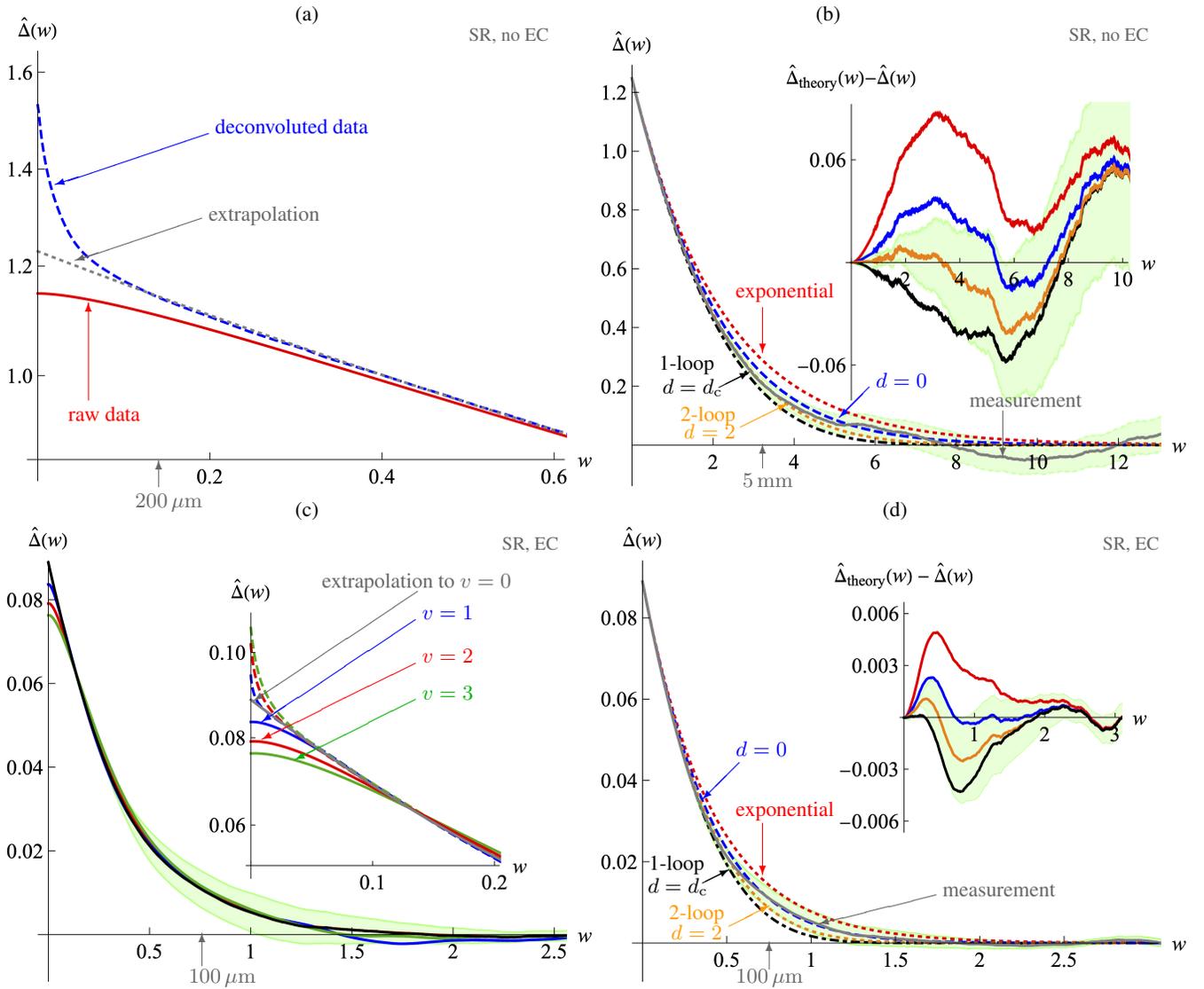}

 We analyse   $\hat\Delta(w)$ in different   materials, summarised in Table~\ref{tab1}. We also consider data where  eddy  currents (EC) play a noticeable effect~\cite{DurinZapperi2006b,PapanikolaouBohnSommerDurinZapperiSethna2011, BohnDurinCorreaMachadoDellaPaceChesmanSommer2018, ZapperiCastellanoColaioriDurin2005}, an aspect experimentally tunable by varying the sample thickness~\cite{ZapperiCastellanoColaioriDurin2005, PapanikolaouBohnSommerDurinZapperiSethna2011, BohnDurinCorreaMachadoDellaPaceChesmanSommer2018}. Details  on  samples are given  in App.~\ref{s:Samples}, and  on  the data analysis in App.~\ref{s:Analysisdetails}, including conversion of our units of $w$ to physical space and time.  

\begin{table}[b]
\begin{tabular}{lcc}
\hline\hline
 {\footnotesize sample} &  {\footnotesize \makecell[c]{interactions /\\ eddy currents}} ~ & {\footnotesize correlation length $\rho$}\\
\hline

\footnotesize \makecell[l]{amorphous   FeSiB film}  &\footnotesize SR / no & \footnotesize 7.5 {\rm  ms}  $\approx$ 495 ${\rm \mu m }$ \\ 

\footnotesize amorphous FeCoB ribbon~ &\footnotesize SR / yes & \footnotesize  0.1 {\rm  s}  $\approx$ 67.5 ${\rm \mu m }$ \\ 

\footnotesize \makecell[l]{polycrystalline   NiFe film}   &\footnotesize LR / no& \footnotesize  12.5 {\rm  ms}  $\approx$ 500 ${\rm \mu m }$ \\

\footnotesize polycrystalline FeSi ribbon & \footnotesize LR / yes & \footnotesize 35 {\rm  ms}  $\approx$ 0.9695 ${\rm \mu m }$ \\
\hline \hline
\end{tabular}
\caption{Short-range (SR) and long-range (LR) samples, with and without eddy currents (ECs).}
\label{tab1}
\end{table}

\smallskip

\noindent
{\bf{SR   interactions  without ECs.}}
 Our first sample is an amorphous  $200$-nm-thick FeSiB film. 
  Fig.\ \ref{Fig2}(a) shows that the raw data for $\hat\Delta(w)$ are
rounded in a boundary layer of size $\delta_w \approx 0.6$, due to the finite driving velocity.  
To   obtain {\color{blue}$\hat\Delta(w)$,}  
we use  \Eq{OperatorUnfolding1-text} with $\tau=0.17$.
\begin{figure*}[t]
{\setlength{\unitlength}{1cm}\begin{picture}(17.8,7.2)
\put(8.9,0){ \includegraphics[width=.49\textwidth]{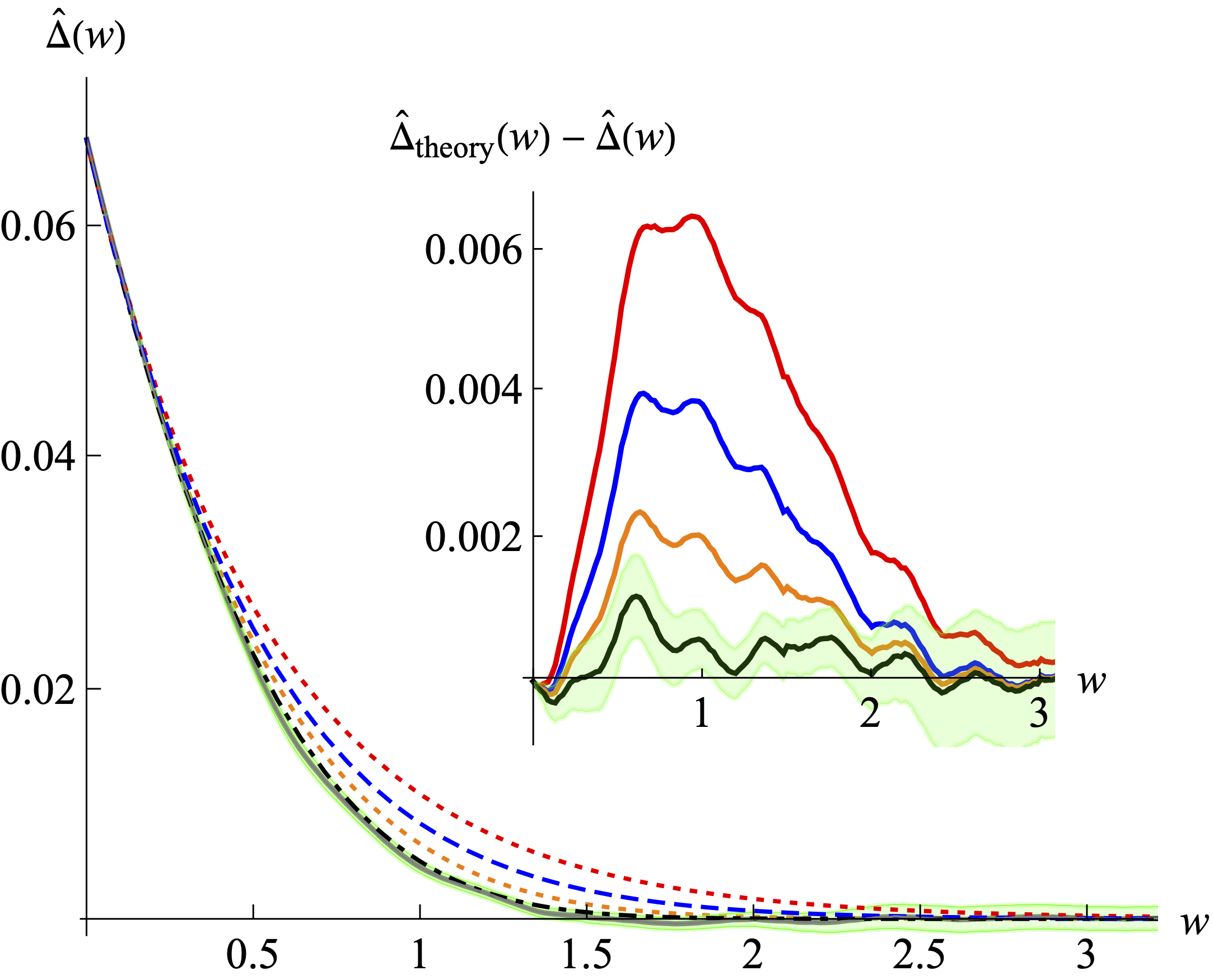}}
\put(0,0){ \includegraphics[width=.49\textwidth]{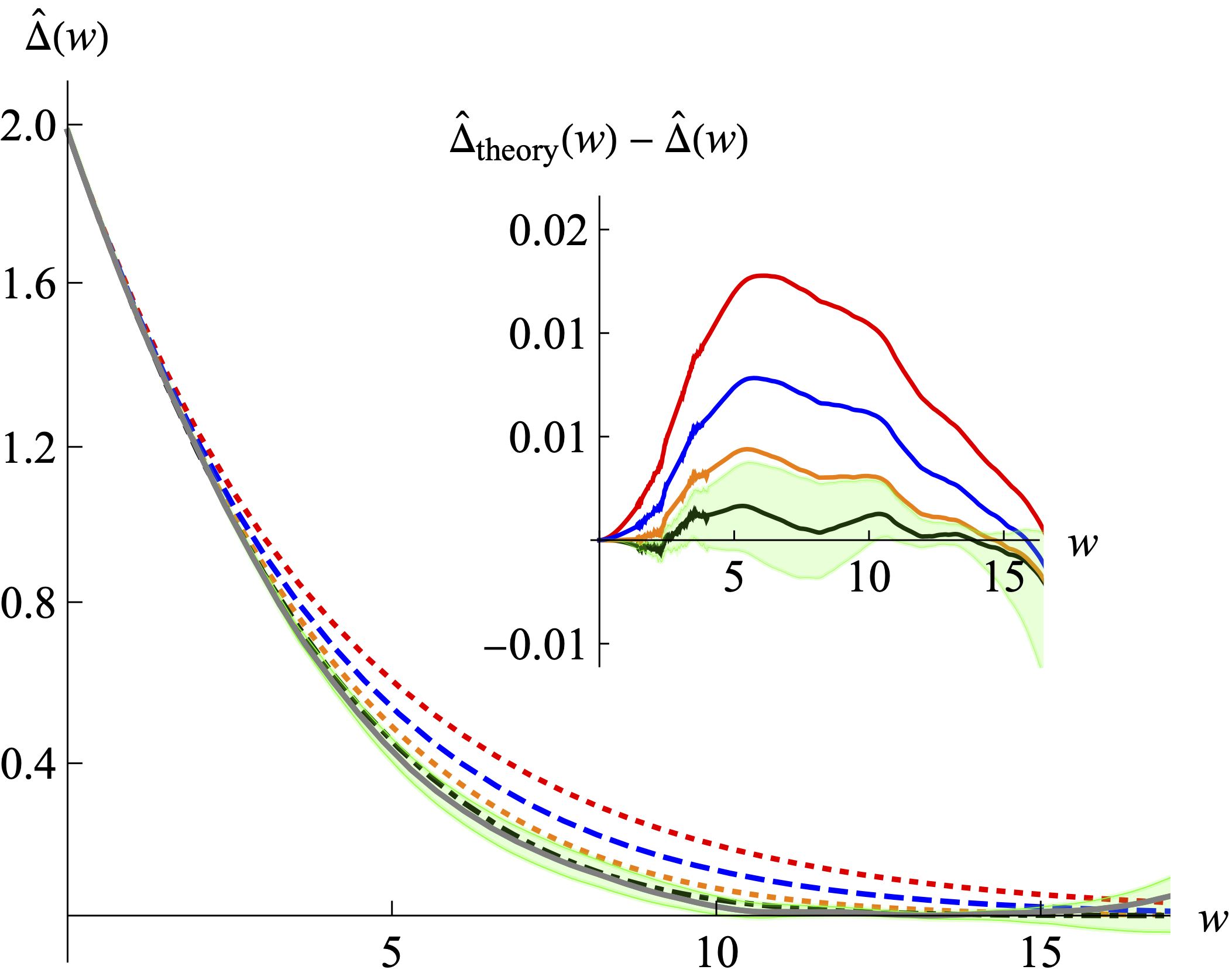}}
\put(4.4,7){(a)}\put(7,6.7){\footnotesize {\color{grey}LR, no EC}}
\put(13.2,7){(b)}\put(16.8,6.7){\footnotesize {\color{grey}LR,   EC}}
\put(10.7,3.4){\color{red} exponential}
\put(11.4,3.35){\color{red}\vector(-0,-1){1.4}}
\put(12.9,1.4){\color{blue} $d=0$}
\put(12.8,1.45){\color{blue}\vector(-1,-1){0.47}}
\put(9.75,1.66){\color{black} 1-loop}
\put(9.75,1.3){\color{black} $d=d_{\rm c}$}
\put(10.7,1.43){\color{black}\vector(2,1){0.53}}
\put(10,0.9){\color{orange} 2-loop}
\put(10.03,0.58){\color{orange} $d=2$}
\put(10.9,0.8){\color{orange}\vector(2,1){0.85}}
\put(13.4,1.1){\color{grey} measurement}
\put(13.9,1.1){\color{grey}\vector(0,-1){0.65}} 
\put(2.4,3.1){\color{red} exponential}
\put(2.63,3){\color{red}\vector(-0,-1){0.55}}
\put(4.3,1.95){\color{blue} $d=0$}
\put(4.2,1.95){\color{blue}\vector(-1,-1){0.55}}
\put(0.95,1.66){\color{black} 1-loop}
\put(0.95,1.3){\color{black} $d=d_{\rm c}$}
\put(1.9,1.4){\color{black}\vector(2,1){0.85}}
\put(2.2,0.9){\color{orange} 2-loop}
\put(2.23,0.58){\color{orange} $d=2$}
\put(3.1,0.75){\color{orange}\vector(2,1){0.75}}
\put(4.75,1.2){\color{grey} measurement}
\put(5.38,1.1){\color{grey}\vector(0,-1){0.56}}
\put(.95,0){\color{grey} $200\,\rm \mu m$}
\put(1.5,0.15){\color{grey}\vector(0,1){.3}}
\put(11.2,0){\color{grey} $1\,\rm \mu m$}
\put(11.4,0.15){\color{grey}\vector(0,1){.3}} 
\end{picture} 
}  
\caption{The measured function $\hat\Delta(w)$ for   our two LR samples: (a) a polycrystalline $200$-nm-thick NiFe film  (negligible ECs), and (b) a polycrystalline FeSi ribbon (with ECs).  They agree  with   1-loop FRG relevant here.  $w=1$  corresponds to $ 2.5 \,{\rm ms} \approx 100\,\mu\rm m$ for (a), and  $ 50 \,{\rm ms}\approx 1.385\,\mu\rm m$ for (b), see grey arrows.}
\label{Fig3}
\end{figure*} 
This reduces the boundary layer (non-straight part) from $\delta_w \approx 0.6$ to $\delta_w \approx 0.1$, allowing us to extrapolate to $w=0$ (grey in Fig.\ \ref{Fig2}(a,b)).
The measured values for $\hat\Delta(0)$ and $\hat\Delta'(0^+)$ are then used to fix all scales in the theory predictions we wish to compare to  on   Fig.~\ref{Fig2}(b). 
These are from bottom to top (analytic expressions are in App.~\ref{s:theory-Delta}):   1-loop FRG (relevant for $d=d_{\rm c}$, i.e.\ LR elasticity), $2$-loop FRG in $d= 2$ (relevant for SR elasticity)~\cite{ChauveLeDoussalWiese2000a,LeDoussalWieseChauve2002}, the $d = 0$ solution~\cite{LeDoussalWiese2008a,terBurgWiese2020} and an exponential, the latter, not realized in magnets, given as reference.  The data agree best, and within error bars,  with the 2-loop FRG prediction for $d = 2$. From   Fig.~\ref{Fig2}(b)  we extract a correlation length $\rho := \hat\Delta(0)/\hat\Delta^\prime(0) \approx 3 $. This agrees with the scale on which $\hat\Delta_{\dot{u}}(w)$ decays to 0 (see   Fig.~\ref{Fig8Suppl}(a)  in App.~\ref{s:Velocity correlations}).

\smallskip

\noindent 
{\bf{SR interactions with ECs.}} Our second sample with SR elasticity is  an amorphous FeCoB ribbon  where ECs are non-negligible. A range of different driving velocities is at our disposal. As  ECs   are  more relevant as $v$ increases, we focus   on    $v = 1,2,3$. 
There  is additional (white) noise contributing to  $\dot{u}$. After integration this   contributes a linear function to $\hat\Delta(w)$, 
s.t.
\begin{align}
\hat\Delta_v^{\rm raw}(0)-\hat\Delta_v^{\rm raw}(w) = \hat\Delta_v(0) - \hat\Delta_v(w) + \sigma_{\rm noise} |w|,  \label{DeltaNoisecontr}
\end{align}
necessitating to subtract a linear term $\sigma_{\rm noise} |w|$ (see  Fig.~\ref{Fig9}  in App.~\ref{s:SRWoutECdetails}).    Fig.~\ref{Fig2}(c) shows $\hat\Delta_v(w)$ after this   subtraction.  
The inset   zooms  into  the boundary layer with  deconvolution by \Eq{OperatorUnfolding1-text}   in the same  color code. Having data at different $v$   allows us to test that 
\begin{enumerate}
\item[(i)] the boundary layer scales linearly in $v$, i.e. $\delta_w\sim v \tau$. 
\item [(ii)]   $\hat\Delta_v(w)$ for $v = 1,2, 3$ unfold to the same   $\hat\Delta(w)$.
\end{enumerate} 
Both conditions are satisfied 
using $ \tau= 0.025$. 
Comparison to the theory proceeds as before, and is shown in   Fig.~\ref{Fig2}(d),  combining $v=1$ and $v=2$ to improve the statistics. Although   error bars are  non-negligible, the data is in agreement with the predicted 2-loop  result in $d= 2$, as for   FeSiB   with SR elasticity  without ECs in  Fig.~\ref{Fig2}(b). 
For $w > 0.7$ the data slightly deviate from the 2-loop result, albeit well within error bars. 
 Either this is   a statistical fluctuation, or due to ECs. 

\smallskip

\noindent 
{\bf{LR   interactions  without ECs.}}
LR elasticity arises in materials, here a   polycrystalline  $200$-nm-thick NiFe film, due to    strong dipolar interactions between parts of the   domain  wall. For long-range elasticity the upper critical dimension $d_{\rm c} = 2$ coincides with the dimension of the  wall. 
The common belief is that then MF theory, i.e.\ the ABBM model, is sufficient to describe the system. 
A glance at  Fig.~\ref{Fig3}(a)  shows that the experimental result is in contradiction to the prediction \eq{DeltaABBM} of   ABBM. While the latter holds at small $w$,   at larger $w$ the correlator $\hat\Delta(w)$ decays to zero. 
Field theory predicts~\cite{ChauveLeDoussalWiese2000a,LeDoussalWieseChauve2002,FedorenkoStepanow2002,LeDoussalWiese2003a} that fluctuations are relevant   at the upper critical dimension, and that    $\hat\Delta(w)$ is given by   1-loop FRG.   Fig.~\ref{Fig3}(a) shows that this is indeed the case.

\smallskip

\noindent
{\bf{LR  interactions  with ECs.}} 
Our fourth sample is a  polycrystalline FeSi ribbon where the elasticity is LR and  ECs  are non-negligible.   Fig.~\ref{Fig3}(b)  shows a comparison of $\hat \Delta(w)$ to the four theory candidates. 
As for the   NiFe film with LR elasticity  and no  ECs, the  agreement is excellent with   1-loop FRG, and inconsistent with ABBM.  We refer to   App.~\ref{s:LRWECdetails}  and Fig.~\ref{Fig10}  for details on the data analysis for this sample.

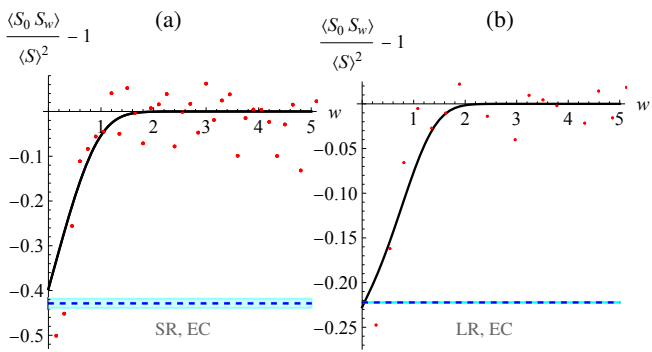
\begin{figure}[t]
\fboxsep0mm
{\setlength{\unitlength}{1cm}\begin{picture}(8.5,4.6)
\put(2,4.3){(a)}
\put(6.4,4.3){(b)}
\put(2,0.2){\color{grey}\scriptsize SR, EC}
\put(6,0.2){\color{grey}\scriptsize LR, EC}
\put(0,0){\includegraphics[width=.25\textwidth]{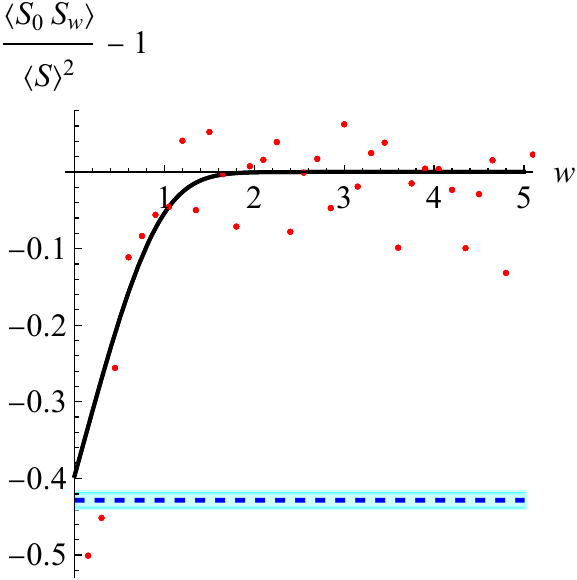}}
\put(4.1,0){\includegraphics[width=.25\textwidth]{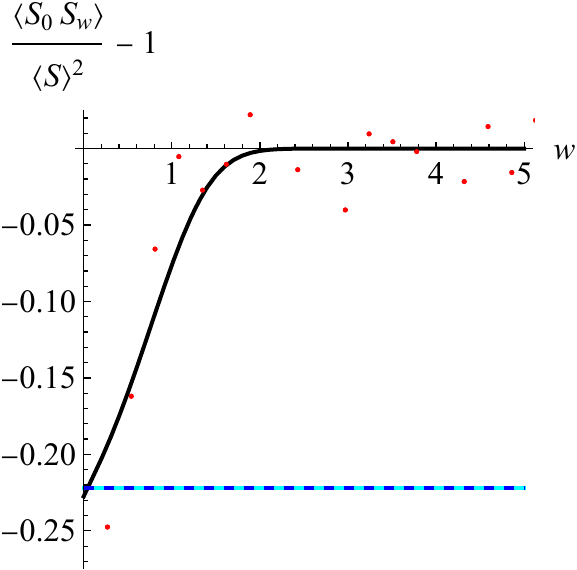}}
\end{picture}
}
\caption{Anticorrelation of avalanches as a function of $w$   as defined in \Eq{e:anticorrelations}, for the two samples with  ECs  (red dots), (a)   FeCoB with SR elasticity and (b)   FeSi   with LR elasticity. The solid line is the prediction $-\hat\Delta''(w)$ of \Eq{e:anticorrelations} from the experiment. The dashed lines are bounds on the maximal  reduction from the field-theory \eq{bound}, with error bars in cyan. $w=1$  corresponds to  $200\, {\rm ms}\approx 135\, \mu\rm m$ for (a) and   $ 50 \,{\rm ms}\approx 1.385\,\mu\rm m$ for (b).}
\label{Fig4}
\end{figure}

In experiments,  force correlations are  ounded, and do not grow indefinitely as in MF models such as   ABBM~\cite{AlessandroBeatriceBertottiMontorsi1990,AlessandroBeatriceBertottiMontorsi1990b,Colaiori2008},  see \Eq{DeltaABBM}. 
As a consequence (Ref.~\cite{Wiese2021} section 4.20, or~\cite{ThieryLeDoussalWiese2016}, Eq.~(8)), avalanches are anti-correlated 
\be\label{e:anticorrelations}
\frac {\left< S_{w_1} S_{w_2} \right>}{\left< S\right> ^2}-1  = - {\hat\Delta''(w_1-w_2)}. 
\ee
Here $S_w$ is the size of an avalanche at $w$, and   $\left<S_w\right> = \left< S\right>$. The numerator   $\left< S_{w_1} S_{w_2} \right> $ is
the expectation of the product of avalanche sizes, given that one is triggered at $w=w_1$, and a second at $w=w_2$;    depends on $|w_1-w_2|$, and is averaged over the remaining variable.
The experimental verification of this relation is shown on Fig.~\ref{Fig4}. Despite   large statistical fluctuations, both the functional form as the amplitude agree. 
Since $\hat\Delta(w)$ is convex, $\hat\Delta''(w)\ge 0$. On the other hand,  $ {\left< S_{w_1} S_{w_2} \right>\ge 0}$, thus $\hat\Delta''(w)\le 1$. This bound is   impossible  to reach, as the toy-model \eq{Delta-DPM} in    $d=0$   has $\hat\Delta''(0^+)=0.5$. 
The field theory~\cite{Wiese2021} gives 
\be\label{bound}
\hat\Delta''(0^+)\le \frac29 + 0.107533 \epsilon + \ca O(\epsilon^2), 
\ee
which evaluates to $0.437$ for   SR ($\epsilon=2$), and $0.222$ ($\epsilon=0$) for   LR correlations.  
 Fig.~\ref{Fig4} shows that this bound is saturated, both for the SR and LR sample. This is surprising as both systems  have multiple domain walls, estimated to be around five for the samples on Fig.~\ref{Fig4}. So either all but one domain wall are pinned, or  these multiple walls are so highly correlated that they effectively behave as a single wall. 
 
In this paper, we measured   the effective force or center-of-mass correlations 
 showing that they have a universal form, predicted by the FRG, both for SR and LR elasticity and mostly independent of ECs. We prove that FRG, an alternative to replica symmetry breaking, correctly models subtle details such as the dependence on dimension and the range of interactions. 
We hope this work inspires the experimental community to look beyond commonly studied observables and beyond MF. Further experimental systems  to explore are sheered colloids or foams, DNA unzipping,
and earthquakes.

\acknowledgements
We thank A.~Douin, F.~Lechenault, G.~Mukerjee and A.~Rosso for discussions. 
  F.B.\ and R.L.S.\ acknowledge financial support from   CNPq and CAPES.

\ifx\doi\undefined
\providecommand{\doi}[2]{\href{http://dx.doi.org/#1}{#2}}
\else
\renewcommand{\doi}[2]{\href{http://dx.doi.org/#1}{#2}}
\fi
\providecommand{\link}[2]{\href{#1}{#2}}
\providecommand{\arxiv}[1]{\href{http://arxiv.org/abs/#1}{#1}}
\providecommand{\hal}[1]{\href{https://hal.archives-ouvertes.fr/hal-#1}{hal-#1}}
\providecommand{\mrnumber}[1]{\href{https://mathscinet.ams.org/mathscinet/search/publdoc.html?pg1=MR&s1=#1&loc=fromreflist}{MR#1}}

\eject

\appendix

\noindent{\bf Supplementary Material for ``Force Correlations in Disordered Magnets'' by C.\ ter Burg, F.\ Bohn, G.~Durin, R.L.\ Sommer, and K.J. Wiese}

\section{Samples  and experiments}\label{s:Samples}
In this work,  we analyze     force correlations in soft magnetic materials. 
We employ two thin films and two ribbons to perform our Barkhausen-noise experiments. The thin films consist of an amorphous Fe$_{75}$Si$_{15}$B$_{10}$ (FeSiB) film and a polycrystalline Ni$_{81}$Fe$_{19}$ (NiFe) film, both with a thickness of $200$~nm. 
The films are prepared by magnetron sputtering onto glass substrates, with dimensions $10$ mm $\times$ $4$ mm, using the parameters   given  in Refs.~\cite{BohnDurinCorreaMachadoDellaPaceChesmanSommer2018}. 
Detailed information on the structural and magnetic characterizations is provided in Refs.~\cite{PapanikolaouBohnSommerDurinZapperiSethna2011,SilvaCorreaPaceCidKernCararaChesmanSantosRodriguez-SuarezAzevedoRezendeBohn2017,BohnDurinCorreaMachadoDellaPaceChesmanSommer2018}. 
Our ribbons   are an amorphous Fe$_{64}$Co$_{21}$B$_{15}$ (FeCoB) and a polycrystalline FeSi alloy with Si=$7.8\%$.
Both ribbons have dimensions of about $20$~cm $\times 1$~cm, with thickness of $\sim 20\,\mu$m.
Further information on the ribbons and their magnetic behavior   are  given in Refs.~\cite{DurinZapperi2000,DurinZapperi2006b}.

Regarding the Barkhausen experiments, we record noise time series using the traditional inductive technique in an open magnetic circuit, in which one detects  
voltage pulses with a pickup coil wound around a ferromagnetic material submitted to a smooth, slow-varying external magnetic field. 
In our setup, sample and pickup coils are inserted in a long solenoid with compensation for the borders to ensure an homogeneous applied magnetic field on the sample. 
The sample is driven by a triangular magnetic field, applied along the main axis of the sample, with an amplitude high enough to saturate it magnetically. 
The  pickup coil is wound around the central part of the sample. 
A second pickup coil, with the same cross section and number of turns, is used to compensate the signal induced by the magnetizing field. 
The Barkhausen signal is then amplified, filtered, and finally digitalized. 

For the thin films, the Barkhausen experiments are performed in Brazil. 
The measurements are carried out using a pickup coil with $400$~turns, $3.5$~mm long and $4.5$~mm wide, and under similar conditions, i.e.,~$50$~mHz triangular magnetic field, $100$ kHz $12$-dB/octave low-pass filter set in the preamplifier (SR$560$ Stanford Research Systems) and signal acquisition taken with an analog-to-digital converter board (PCI-DAS$4020/12$ Measurement Computing) with sampling rate of $4\times 10^6$ samples per second~\cite{BohnDurinCorreaMachadoDellaPaceChesmanSommer2018}. 
At a preanalysis stage, we   employ a Wiener deconvolution~\cite{PapanikolaouBohnSommerDurinZapperiSethna2011}, which optimally filters the background noise and removes distortions introduced by the response functions of the measurement apparatus in the original voltage pulses, thus providing reliable statistics despite the reduced intensity of the signal. 

For the ribbons, the experiments are performed in Italy. 
They are carried out using a pickup coil with $50$ turns, $1$~mm long and $1$~cm wide, a triangular magnetic field with frequency between $3$-$50$~mHz, and a low-pass preamplifier filter chosen in the $3$-$20$~kHz range, roughly half of the sampling rate.   Specifically, we consider sampling rate of $50\times 10^3$~samples per second for FeCoB, and $20\times 10^3$~samples per second for FeSi~\cite{DurinZapperi2000,DurinZapperi2006b}. 
For the FeCoB ribbon, the sample is submitted to a small tensile stress of $2$~MPa during the measurement in order to enhance the signal-to-noise ratio. 

All   time series for films and ribbons are acquired   around the central part of the hysteresis loop, near the coercive field, where the domain wall motion is the main magnetization mechanism and the noise achieves the condition of stationarity~\cite{DurinZapperi2006b}. 
For each experimental run, the statistical properties are obtained from at least $150$ measured   time series.

While the central issue in this work  is to explore the force correlations from the Barkhausen-noise time series, 
the classification into the different universality classes reposes on earlier work, 
where we identified the universality class of Barkhausen avalanches by measuring the distributions of   avalanche sizes and   durations, the average size as a function of the avalanche duration, their power spectrum, and the average avalanche shape. 
The results for the thin films  can be found in  Refs.~\cite{PapanikolaouBohnSommerDurinZapperiSethna2011, BohnDurinCorreaMachadoDellaPaceChesmanSommer2018},  
 the ones for the ribbons  in Refs.~\cite{DurinZapperi2000,DurinZapperi2006b}.

\section{Subtraction of the baseline, measurement of $\hat\Delta(w)$,  and error estimates}
\label{s:Numerical methods}

\subsection{Correcting the baseline}
 
\begin{figure}[t]
\fboxsep0mm
\setlength{\unitlength}{1cm}
{\begin{picture}(8.6,6.0)
\put(0,0){\includegraphics[width=8.6cm]{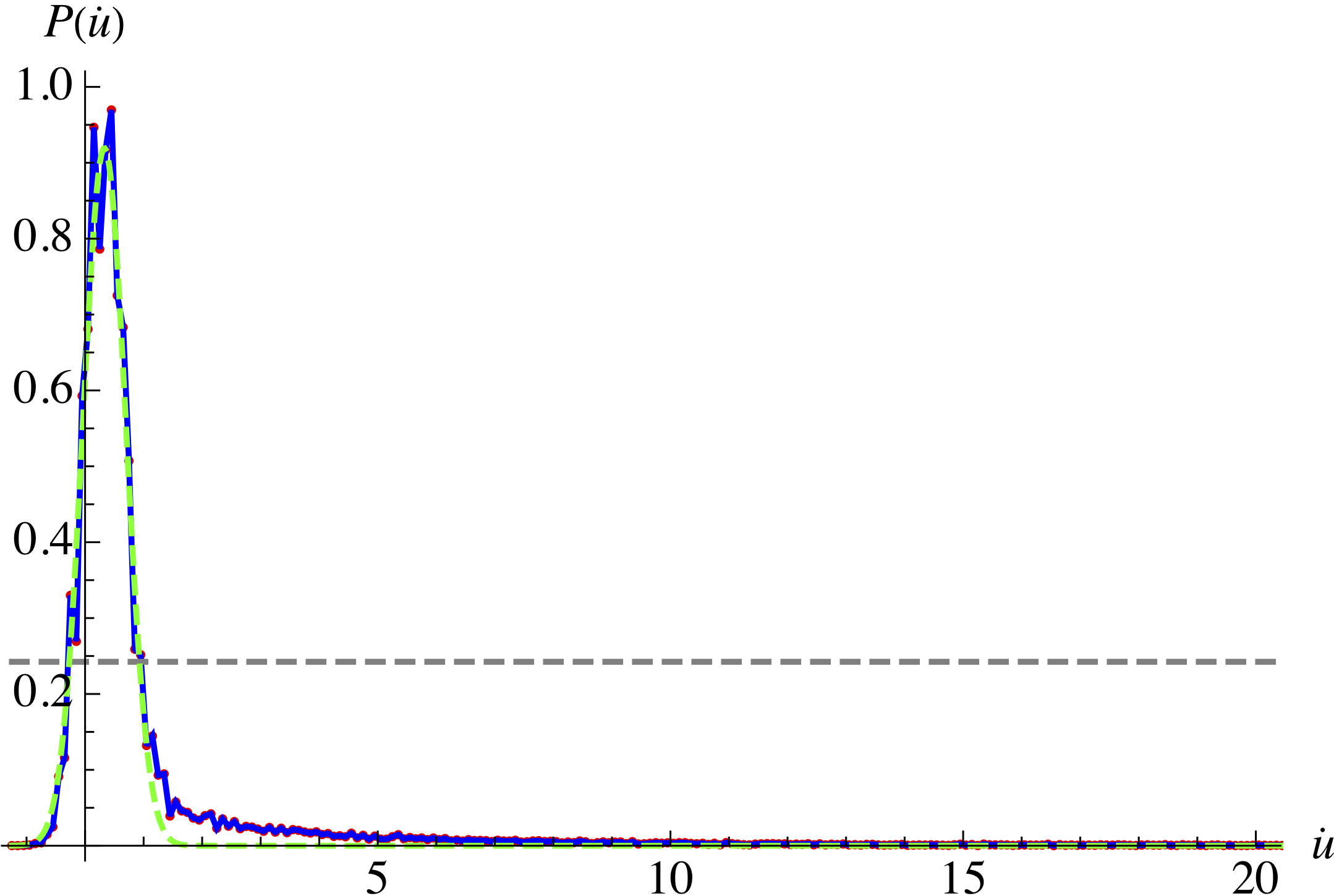}}
\put(2.2,1.8){\includegraphics[width=6.3cm]{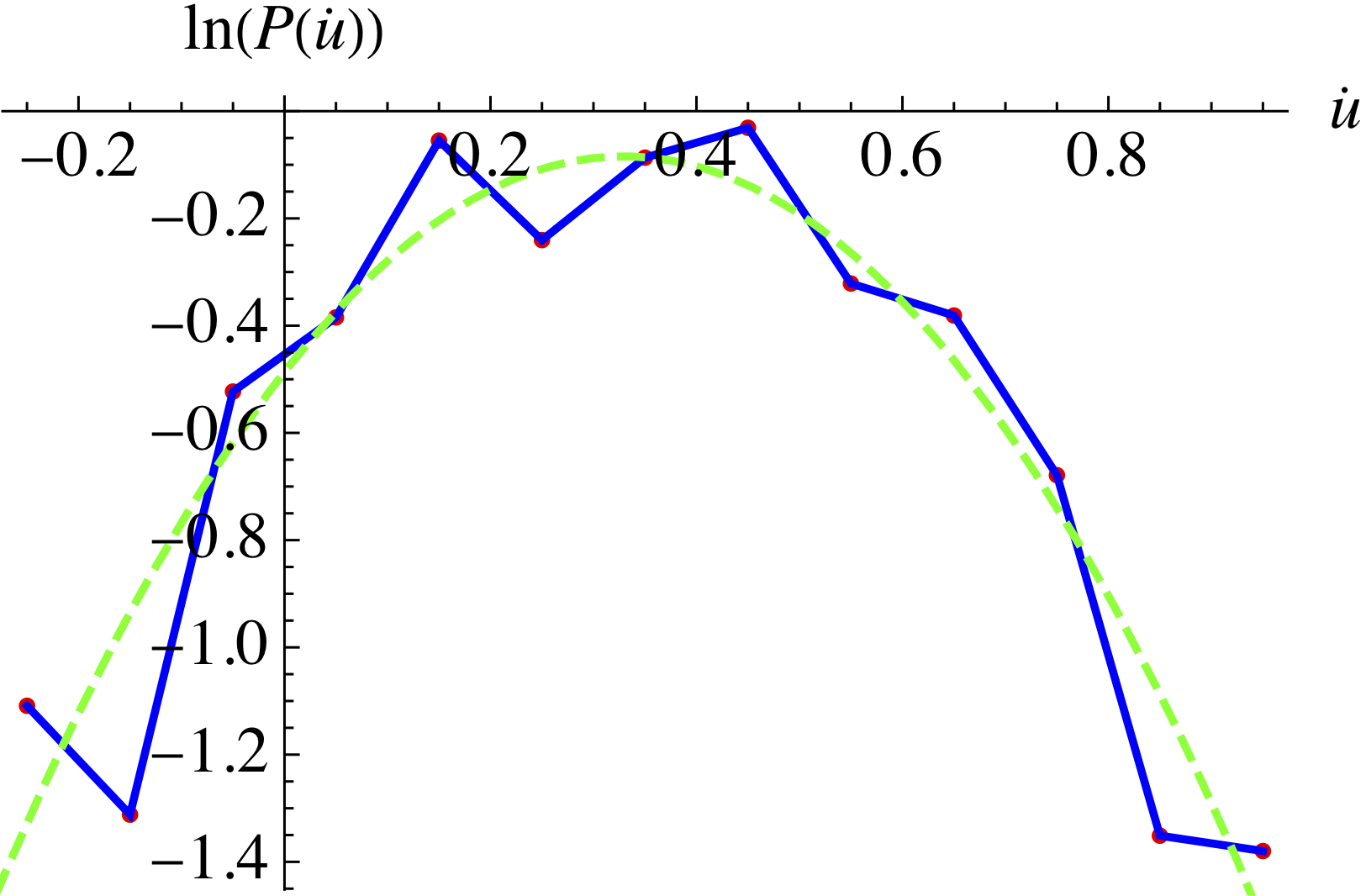}}
\end{picture}}
\caption{Distribution $P(\dot{u})$ (blue, solid) with fit (green, dashed) to all data points above the dashed line. It is obtained from the optimal parabolic fit for $\ln P(\dot u)$, as shown in the inset. The $\dot u$-value at the maximum of the fit is used as the   position for the zero of $\dot u$.} 
\label{P-of-udot}
\end{figure} 
Here we present the methods used to obtain the   correlator $\hat\Delta_v(w)$ defined in \Eq{obs} from the experimental data for the change in flux $\dot u_{w=vt} \equiv \dot{u}(t)$. 
As the magnetic field is increased at a rate $v$ 
\begin{align}
u_w -w = \int_0^{w/v} {\rm d}t\, [ \dot{u}^{\rm raw}(t) - v ]  . \label{FwIntegrated}
\end{align}
We found that there are strong {\em run-to-run} fluctuations for the mean $v_i :=   \langle \dot{u}_{\rm raw} \rangle_i$
in run $i$,  due to a drift in the amplifier baseline. 
If the estimate for $v$ in \eqref{FwIntegrated} is not correct, 
this adds a term of the form $c w$ to $u_w-w$, with $c$ a random number. If we suppose that $c$ is Gauss-distributed with mean $0$, 
    integration   leads to a parabolic   contribution, i.e.\ $\hat\Delta_v(w)\to \hat\Delta_v(w) + \half \left< c^2\right> w^2$. To correct   this, we  proceed as  follows: For each  run $i$ we consider the distribution $  P(\dot{u})$ (see Fig.~\ref{P-of-udot}), and fit a Gaussian to its peak. This is done by choosing the  data points which satisfy $P(\dot u)>0.25 \max_{\dot u}P(\dot u)$, and then fitting a parabola to  $\ln [P(\dot u)]$. Finally, $\dot u$ is shifted so that the maximum of the parabola lies at $\dot u=0$. 
Our best estimate for the driving velocity is then the  average over $N$ runs
\begin{align}
v = \frac{1}{N} \sum_{i = 1}^{N} \langle \dot{u} \rangle_i. 
\end{align}
 This allows us   to construct the interface position $u_w-w$ for run $i$ as
\begin{align}
 u^i_w - w = \int_0^{w/v} \rmd t\, [ \dot{u}_i(t) - v ]  . \label{FwIntegratedCorr}
\end{align}
The experimental setup makes appear an additional numerical prefactor on the r.h.s.\ of \Eq{FwIntegratedCorr}. It is eliminated by demanding that the linearly increasing parts of Fig.~\ref{FigSRNECForcedrops}   have slope~$1$.

\subsection{Error bars}
\label{s:error-bars}

\begin{figure}[t]
\includegraphics[width=1\columnwidth]{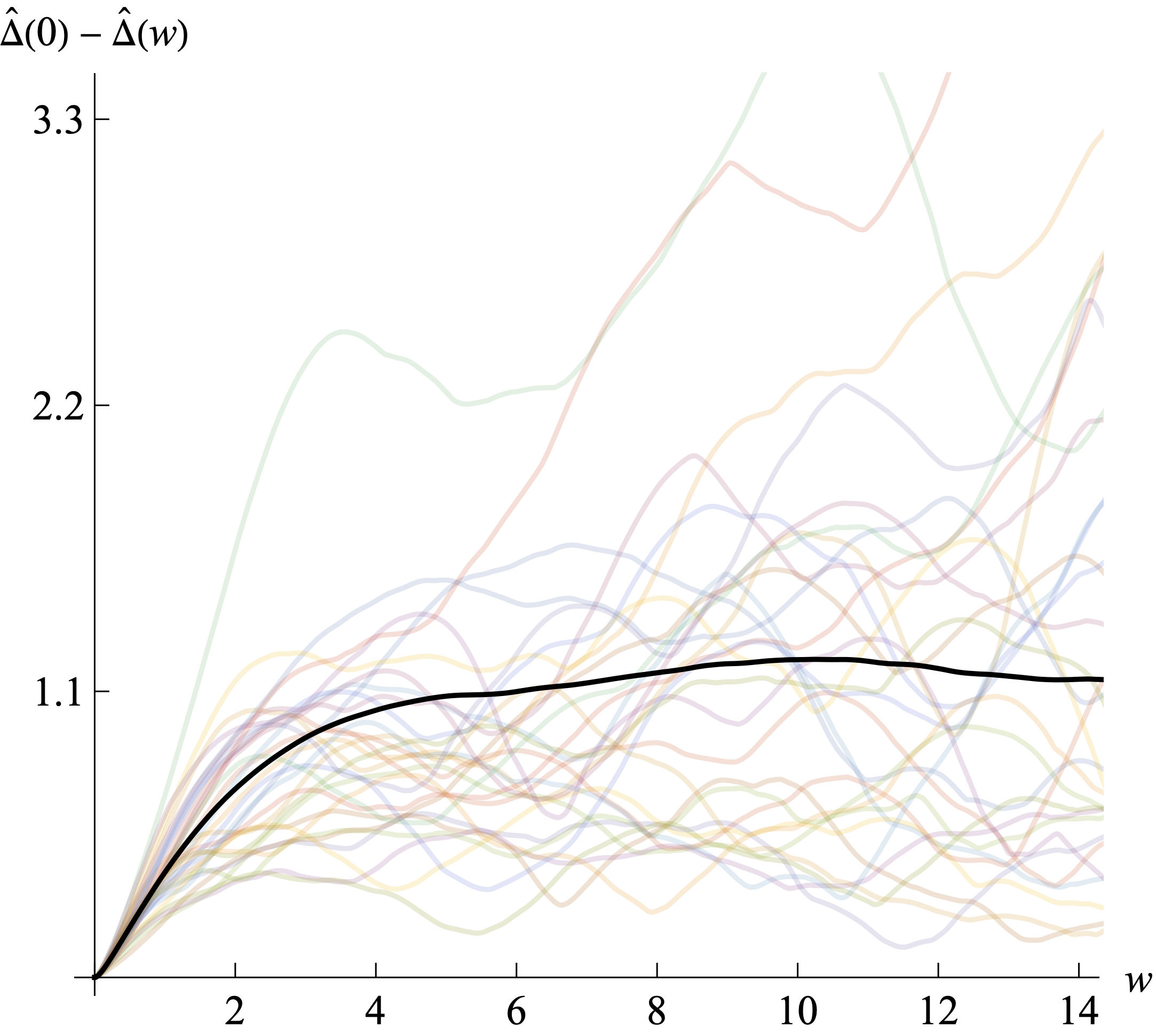} 
\caption{The correlator $\hat\Delta_v(w)$ as given by \Eq{DefDelta} for   the FeSiB film (black) with SR elasticity  and no ECs.  Shaded in the background are the averages $\ca M_i$ over a single sweep  from \eqref{SepMi} which show  strong run-to-run fluctuations.}  
\label{Fig7Suppl}
\end{figure}%
 The  connected two-point correlations of the center of mass for run $i$ are 
\begin{align}
\mathcal{M}_i (w - w^\prime) := \frac{1}{2}\left< \left[(u_w-w) -(u_{w'}-w') \right]^2 \right>_i .   \label{SepMi}
\end{align}
Whereas the   $\mathcal{M}_i$ show strong fluctuations (see Fig.~\ref{Fig7Suppl}), their mean (black) 
\begin{align}
\hat\Delta_v(0) - \hat\Delta_v(w) = \frac{1}{N} \sum_{i  = 1}^N \mathcal{M}_i (w )  , \label{DefDelta}
\end{align} 
is much more stable. Statistically, the small-$w$ region is more robust than the large-$w$ tail. 

Given $\ca M_i(w)$, we need to achieve two goals: (i) extract the plateau height $\hat \Delta(0) = \lim _{w\to \infty} \hat \Delta(0)- \hat \Delta(w)$ for large $w$ (see  Fig.~\ref{Fig7Suppl}), and (ii) estimate the statistical error.  
Due to the large fluctuations between   runs,    the distribution of  $\mathcal{M}_i$   is not a Gaussian, and standard tools for error analysis fail. The key to solve this problem is to observe that the central limit theorem still applies:
partial means over $n<N$ runs have a statistics which increasing $n$ gets closer and closer to that of a Gaussian. As we have always at least  $N=150$ runs, this improvement is substantial, as we can take $n$ up to $n=N/2$. This procedure is known as the {\em  statistical resampling method}:
 One randomly divides all datasets into two parts and computes the variance of the partial means. Averaging this over 100 random partitions gives a   robust estimate for the variance. 
This is formalized in  appendix A of~\cite{WieseBercyMelkonyanBizebard2019}. 
To obtain the error bars for the shape shown in the main text, all partial means have been rescaled such that their derivative at $w = 0$ equals the mean of $\hat\Delta^\prime (0^+)$ over all runs. 
Only then statistical resampling is applied. 
This takes out amplitude fluctuations and reduces the error bars to   errors of the shape. 
To summarize,  our experiment  
for given $w$ can be modeled as 
\be\label{B6}
\hat \Delta_{\rm exp }(w) =\hat \Delta (w) + \sigma(w) \eta, 
\ee
where $\eta$ is a Guassian random variable with mean zero and variance $1$.


\subsection{Estimate of the total error}
\label{s:theorycheck}

In section \ref{s:error-bars} we   obtained  error-bars of the shape $\hat \Delta(w)$. We still need to put a number on how large the deviations of the measured $\hat \Delta(w)$ and the theory are. If \Eq{B6} holds and  the (a priori unknown) $\hat \Delta(w) = \hat \Delta_{\rm th}(w)$, then  we can  define a measured  $\eta$ via   
\be\label{B7}
\eta := \frac{\hat \Delta_{\rm exp}(w) - \hat \Delta_{\rm th}(w)}{\sigma(w)}.
\ee
We can turn this into a test: Using $\eta$ defined by \Eq{B7}, it should have variance  $\left< \eta^2\right>\approx1$. If we  measure a (much) larger $\left< \eta^2\right>$, then $\eta$ has a mean, and theory and experiment do not agree. 

The problem of this procedure is that for a given $w$, we only have one sample. Our statistics can be improved by taking the joint measure for all $w$. However, 
the measured values of $\hat \Delta(w)$ are  correlated, and we cannot simply add up their variances. 
We propose   the following global error estimator
\be
\sigma^2 =   \frac1{\rho}  \int_{0}^{w_{\rm max}} \rmd w\,\frac{\big[\hat \Delta_{\rm th} (w) - \hat\Delta_{\rm exp}(w) \big]^2 }{{\sigma(w)} ^2}   .  \label{L2norm}
\ee
This is best thought of as a discrete sum over all $w$, divided by the correlation length in the same discretized units. Stated differently, this is the sum of the mean variances per correlation-length  segments, equivalent to demanding that the function be satisfied {\em simultaneously in each} of the $w_{\rm max}/\rho$ independent segments.

Let us stress that this is the best we can do, and that the variance of the true error may differ by a   numerical factor, such as $0.5$, $2$ or $3$. 
 Values for this expression, or more precisely its square root $\sigma$ are given in table \ref{tab3}. 
  One sees that for the SR samples the agreement is best with the resummed 2-loop FRG. 
For LR samples the best agreement is  with the 1-loop FRG.  We also see that the measured variances  $\sigma^2$  for the best matching theory are in general about four times smaller than the next best one. 
  This corroborates our statements made in the main text.

\begin{table}[t]
\footnotesize
\begin{tabular}{lcccc}
\hline\hline
 $\sigma$  &   {SR No EC}  & { SR EC} &  {  {LR No EC}} ~ & { LR EC} \\
\hline
1-loop FRG   &  1.07$ $&  1.24 & 0.28  &  0.56  \\ 
 2-loop FRG & 0.45 &   0.77 & 0.76 &  1.11 \\ 
$d=0 $   & 1.89 &   0.96 & 2.85 & 2.03   \\
 exponential &  3.20 & 2.95 & 1.55 &  3.83   \\
\hline \hline
\end{tabular}
\caption{Estimation of $\sigma$ defined in \Eq{L2norm} for all combinations of theory  and experimental data.}
\label{tab3}
\end{table}

\section{Theory predictions for the different classes}\label{s:theory-Delta}
 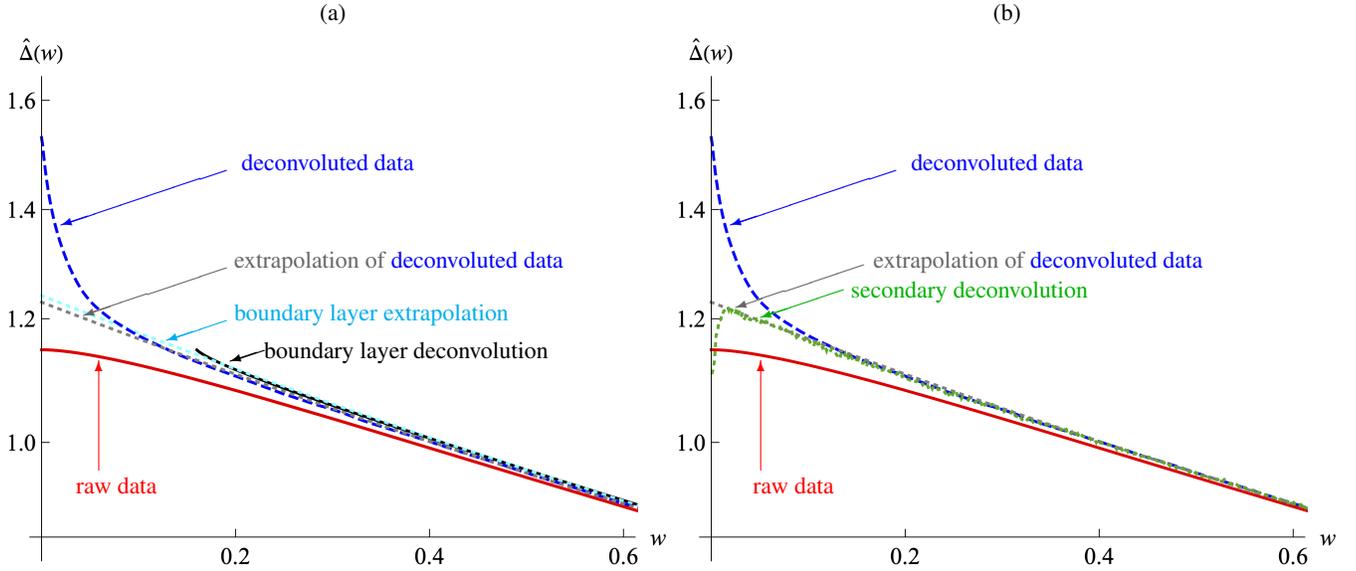
\begin{figure*}[t]
\centerline{(a)\hspace{.48\textwidth}(b)}
{\setlength{\unitlength}{1cm}\begin{picture}(17.8,7.2)
\put(0,0){ \includegraphics[width=.49\textwidth]{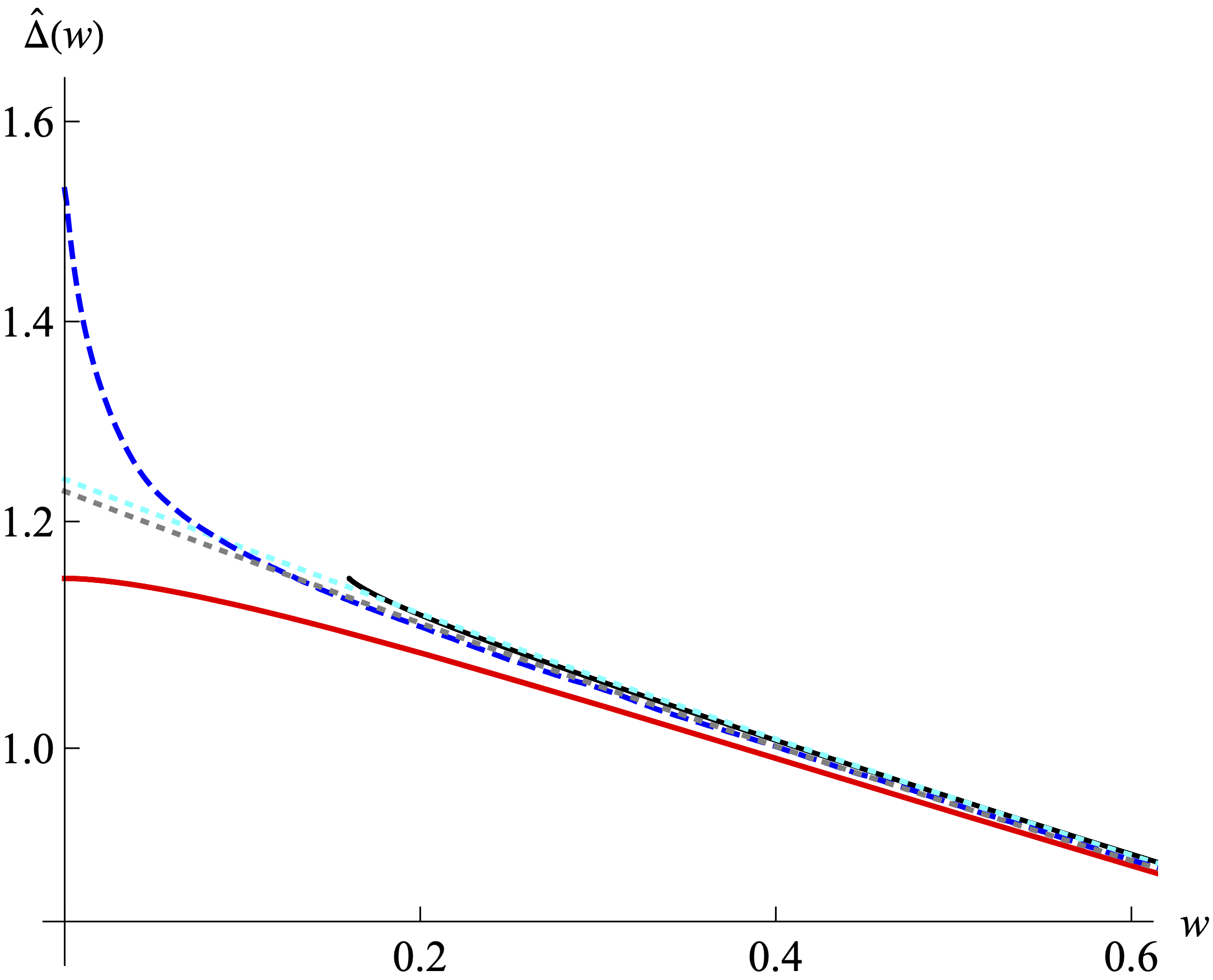}}
\put(8.9,0){ \includegraphics[width=.49\textwidth]{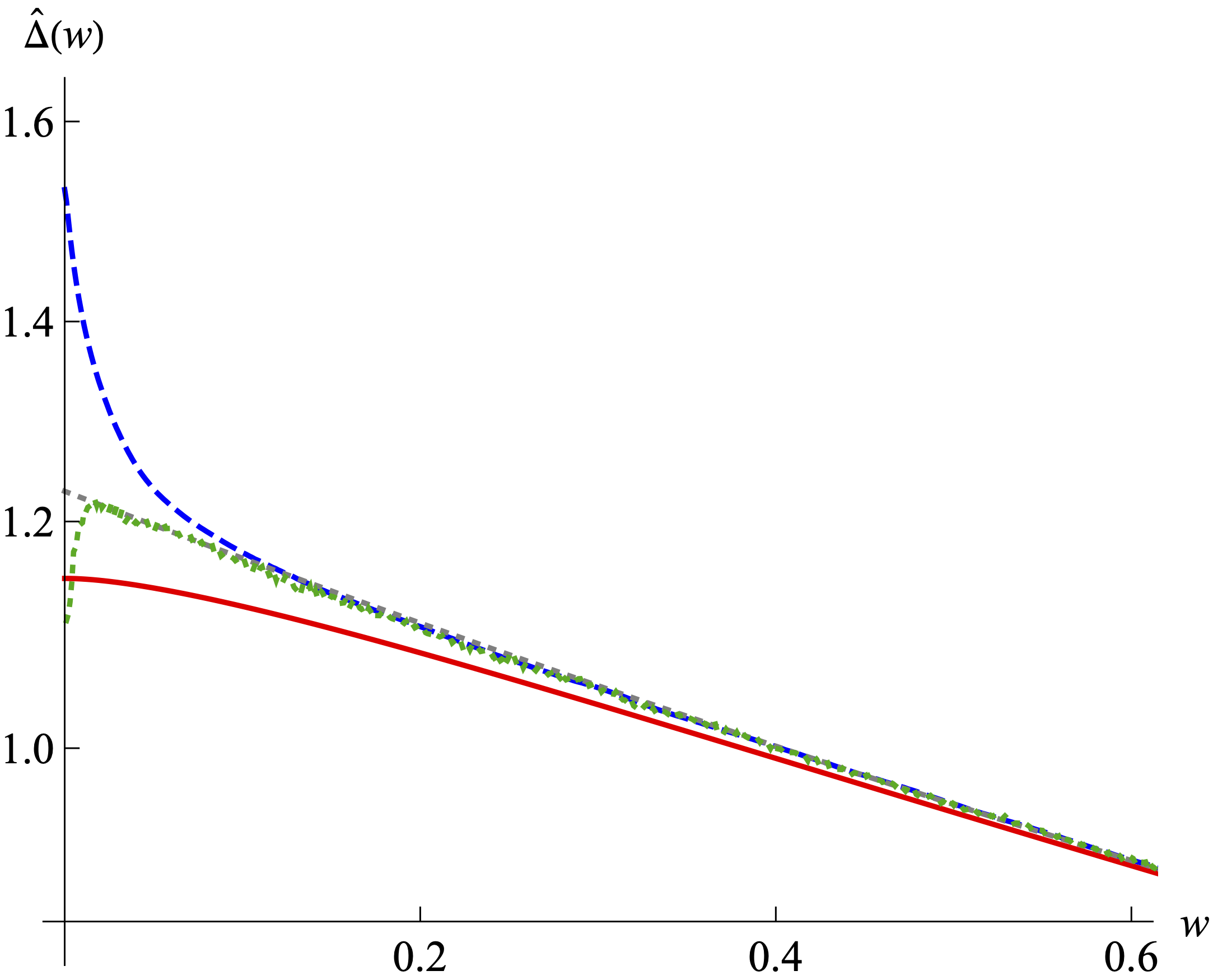}}
\put(1,1){\red raw data}
\put(1.3,1.3){\red\vector(0,5){1.47}}
\put(3.2,5.3){\color{blue} deconvoluted data}
\put(3.0,5.3){\color{blue}\vector(-3,-1){2.2}}
\put(3.1,4.){\color{grey} extrapolation of \color{blue} deconvoluted data}
\put(3,4.){\color{grey}\vector(-3,-1){1.9}}
\put(3.1,3.3){\color{cyan} boundary layer extrapolation}
\put(3,3.3){\color{cyan}\vector(-3,-1){0.8}}
\put(3.5,2.8){\color{black} boundary layer deconvolution}
\put(3.5,2.9){\color{black}\vector(-3,-1){0.45}}
\put(10,1){\red raw data}
\put(10.1,1.3){\red\vector(0,5){1.47}}
\put(12.1,5.3){\color{blue} deconvoluted data}
\put(11.9,5.3){\color{blue}\vector(-3,-1){2.2}}
\put(11.6,4.0){\color{grey} extrapolation of \color{blue} deconvoluted data}
\put(11.47,4.04){\color{grey}\vector(-3,-1){1.7}}
\put(11.3,3.6){\color{darkergreen} secondary deconvolution}
\put(11.2,3.7){\color{darkergreen}\vector(-3,-1){1.1}}
\end{picture}}
\caption{Comparison of three different deconvolution procedure  for   the  $200$-nm-thick FeSiB film with SR elasticity and no ECs  (raw data in solid, red):   deconvolution via \Eq{boundaryLayer} as discussed in the  main text (blue dashed),  time scale    $\tau =0.175$, and extrapolation (grey, dotted). 
In (a) is shown  in addition   deconvolution via the  boundary layer given by \Eqs{boundaryLayer}-\eq{boundaryLayer2} (black, solid)   using the same  $\tau =   0.175$, and its extrapolation (cyan, dotted). 
In (b) is shown the result of       secondary deconvolution using \Eq{OperatorUnfolding2} (green, dotted),    with   $\tau =0. 175$ and  $\tau^\prime =0.00 24$. }
\label{Fig2unfapp}
\end{figure*}%

The extracted fixed-point function $\hat\Delta(w)$ can be compared to   results from the functional renormalization group (FRG) sketched below. 
As we cannot give more than a short summary here, we refer  to sections 2 and 3 of the recent review~\cite{Wiese2021}
for a pedagogic introduction. 

 Contrary to conventional RG, where one considers the flow of a single coupling constant, the FRG follows the flow of an entire function, here the disorder-force correlator $\Delta(w)$ introduced in \Eq{obs}. Writing $\epsilon = d_{\rm c} - d$ for the expansion parameter around the upper critical dimension $d_{\rm c}$, the FRG fixed-point equation for the rescaled (dimensionless) correlator $\tilde \Delta(w)$ reads at 1-loop order (leading order in $\epsilon$)  
\begin{align}
\nn
\partial_l \tilde{\Delta}(w)  \ =  & (\epsilon - 2 \zeta)  \tilde{\Delta}(w) + \zeta u  \tilde{\Delta}^\prime (w) \\ 
\ - &  \frac{\rm d^2}{{\rm d}w^2} \frac{1}{2} \bigl[\tilde{\Delta}(w) - \tilde{\Delta}(0) \bigl]^2 + \dots  
\label{FRGEquation}
\end{align}with the dots representing higher-order contributions~\cite{ChauveLeDoussalWiese2000a,LeDoussalWieseChauve2002} in an expansion in $\epsilon$.   This equation has solutions which decay at least exponentially fast only for selected values of the roughness exponent $\zeta$. For the RF disorder present in the experiment, the appropriate solution of the fixed-point equation $\partial_\ell \tilde \Delta(w) =0$ associated to Eq.~\eqref{FRGEquation} is
\begin{table}[t]
\begin{tabular}{lcc}
\hline\hline
\footnotesize $\bm{\hat\Delta(w)}$  & \footnotesize $\bm{\frac{\hat\Delta(0)\hat\Delta^{\prime \prime }(0)}{\hat\Delta^\prime (0^+)^2}}$ 
& $\hat\Delta''(0)$ \\
\hline
  Exponential~ &   1 & $1$  \\ 

  $d=0$, \Eq{Delta-DPM} & $0.822$ & $\frac 12$  \\
  
   2-loop FRG for $d=2$ (SR)  &   $0.75 $ & $\le 0.437$ \\
   
    1-loop FRG,~\Eq{FP-1loop} (LR) &   $\frac23$ & $\le \frac29$  \\

  SR elasticity without ECs  &   $0.73(3)$ & $0.37(2)$  \\
  
   SR elasticity with ECs  &  $0.65 (10) $ & $0.41(2)$  \\
   
   LR elasticity without ECs  &    $0.58 (10)$ & $0.17(5)$  \\
   
   LR elasticity with ECs  &   $0.65 (10)$ & $0.24(4)$  \\
\hline \hline
\end{tabular}
\caption{Comparison of theoretically and experimentally obtained amplitudes and amplitude ratios.   Theory values are obtained by Taylor-expanding \Eqs{FP-1loop} (1-loop) and \eq{Delta-DPM} ($d=0$). The 2-loop results can be found in~\cite{LeDoussalWieseChauve2002,Wiese2021}.}
\label{tab2}
\end{table}
(see e.g.\ Ref.~\cite{Wiese2021})
\bea \label{FP-1loop}
\tilde{\Delta}^{\mbox{\scriptsize 1-loop}}(w) &=&  -\frac{\epsilon}3 W\biggl(- \exp\bigl(  -\frac{w^2}{2} - 1 \bigl) \biggl)+ \ca O(\epsilon^2),\qquad \\
  \zeta &=&   \frac \epsilon 3+ \ca O(\epsilon^2).
\label{C3}
\eea
Here $W(z)$ is the product-log function, the principle solution for $x$ in $z = xe^x$.
The observable in \Eq{obs} is  obtained from the (scale-free) fixed-point solution $\tilde \Delta(w)$ as 
\bea
\hat \Delta(w)  &=&  \ca A  \hat\rho^{2} \tilde{\Delta}(w /\hat \rho)  \label{FRGsolution},\\ 
 \hat \rho &:=&   \rho \frac{|\tilde \Delta'(0)|}{\tilde \Delta(0)}.
\eea
The amplitude   $\ca A$ is a number (depending on $mL$), whereas the correlation length $\rho$ of  \Eqs{RhoDef} and \eq{FRGsolution} scales as  $\rho\sim m^{- \zeta}$. 
The fixed point \eq{FP-1loop}-\eq{C3} gets corrected at 2-loop order~\cite{ChauveLeDoussalWiese2000a,LeDoussalWieseChauve2002}   as more terms appear in \Eq{FRGEquation}. 
In principle, it allows us to predict $\hat\Delta(w)$ for domain-wall dimensions between $d=4$   down to  $d=0$. 
The bulk magnets used here have $d=2$ ($\epsilon=2$), whereas a thin magnetic film has $d=1$ ($\epsilon=3$). Dimension $d=0$ ($\epsilon=4$) is realized in the DNA/RNA peeling experiment of Ref.~\cite{WieseBercyMelkonyanBizebard2019}. The precision of the approximation decreases with $d$, since the expansion parameter $\epsilon=4-d$ increases. %
\begin{figure}[t]
\includegraphics[width=1\columnwidth]{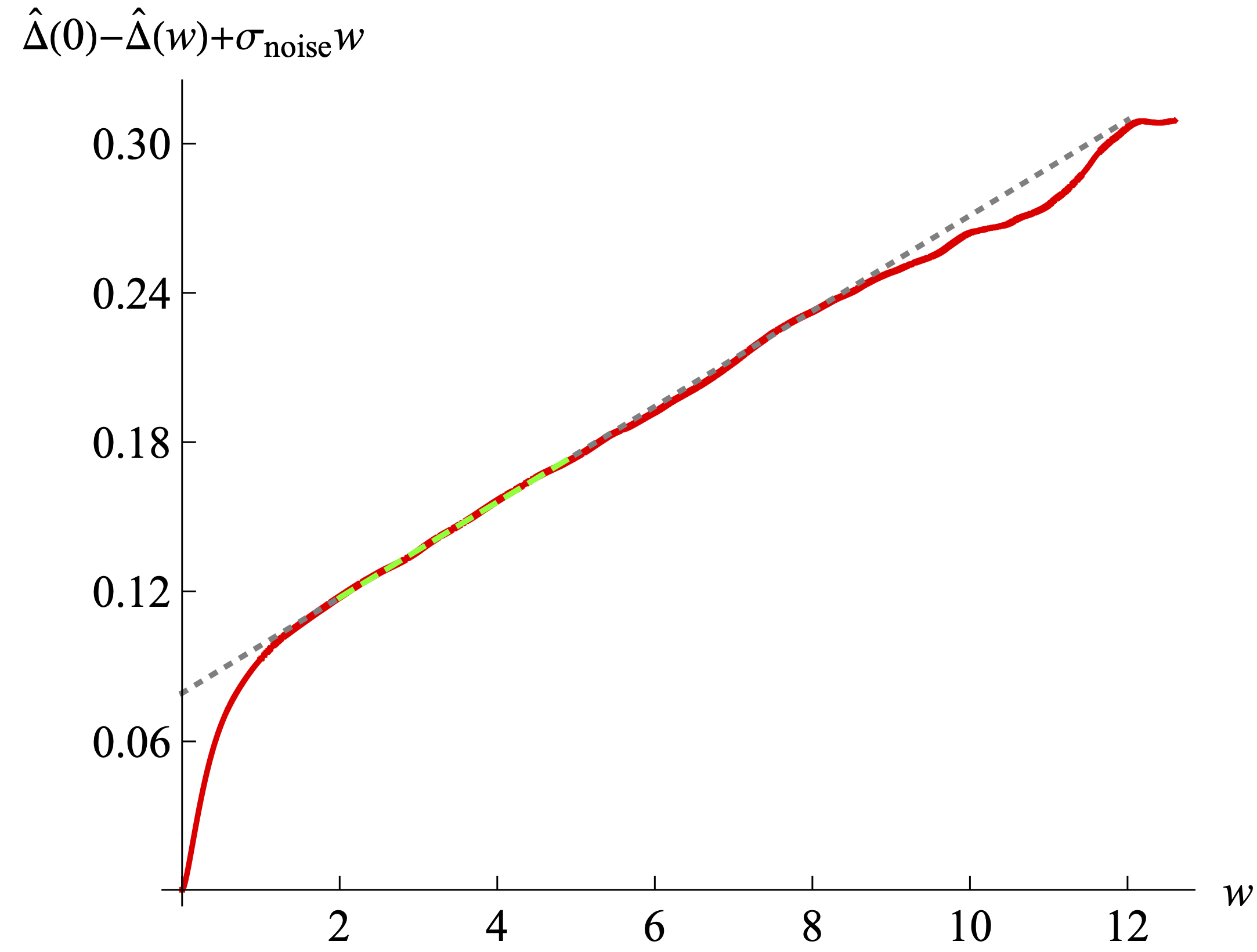}\caption{Subtraction of a linear noise contribution (grey dotted, fitting region in green dashed) $\Delta(0) + \sigma_{\rm noise} |w|$ for   the FeCoB ribbon  with SR elasticity and ECs at $v = 2$. }
\label{Fig9}
\end{figure}%
We are in the fortunate position to have an analytic solution in $d=0$~\cite{LeDoussalWiese2008a,terBurgWiese2020}, 
\be\label{Delta-DPM}
\tilde \Delta^{d=0} (w) =\frac{w^2}2+\mbox{Li}_2(1-\rme^{w})+\frac{\pi^2}6.
\ee
This allows us to choose a Pad\'e approximant for the 2-loop result, optimized for agreement with the   solution \eq{Delta-DPM}.
A summary of properties for $\hat \Delta(w)$ for the different classes is presented in table \ref{tab2}.

%
%
 In the experiment, each function contains two scales, 
 the amplitude $\hat\Delta(0)$  and a correlation length $\rho$ in the $w$-direction. The latter is defined by
\be
\rho : = \frac{\hat\Delta(0)}{|\hat\Delta'(0^+)|}. \label{RhoDef}
\ee
It enters into the scaling form \eq{FRGsolution} as indicated. 
Rescaling the theory candidates to have the same $\hat\Delta(0)$ and $\Delta'(0^+)$ 
 ensures that one compares   the shape without any  fitting parameter.   This is the form used  in the main text. 

 Finally let us mention that  the normalization in \Eq{obs} is different from the one used in the field theory~\cite{LeDoussalWiese2006a}, which contains an additional factor of $m^4L^d$,  with $L$ the system size, on the r.h.s. Our choice is motivated by a lack in the knowledge of  $m^2$ and $L$, and  by the reduction of scales in $\hat\Delta_v(w)$ to a single one, namely the correlation length $\rho:= \hat \Delta(0)/\hat \Delta'(0^+)$ in the $w$-direction.

\section{Deconvolution of $\hat\Delta_v(w)$.}
\label{s:deconvolution}
Suppose the response function decays  exponentially with time scale $\tau$,  
\begin{align}
R(t) = \frac{1 }{\tau} e^{-t/\tau} \Theta(t) .
\label{SimpleResponseFunction}
\end{align}
Then it satisfies the   differential equation 
\begin{align}
 (1 + \tau \partial_t) R(t) = \delta (t) ,\label{ResponseFunctionKernel}
\end{align}
and is normalized, 
\be
\int_{0}^\infty R(t) \rmd t= 1.
\ee
This allows us  to invert Eq.~\eqref{DeltaDefRounded-text} as~\cite{terBurgWiese2020}
\begin{eqnarray}
\hat\Delta(w)  &= & (1 + v \tau \partial_w)(1 - v \tau \partial_w)\hat\Delta_v(w) \nn \\ 
& = &  (1 - (v \tau)^2 \partial^2_w)\hat\Delta_v(w). 
\label{OperatorUnfolding1deltau}
\end{eqnarray}%
\begin{figure}[t]
    \includegraphics[width=1\columnwidth]{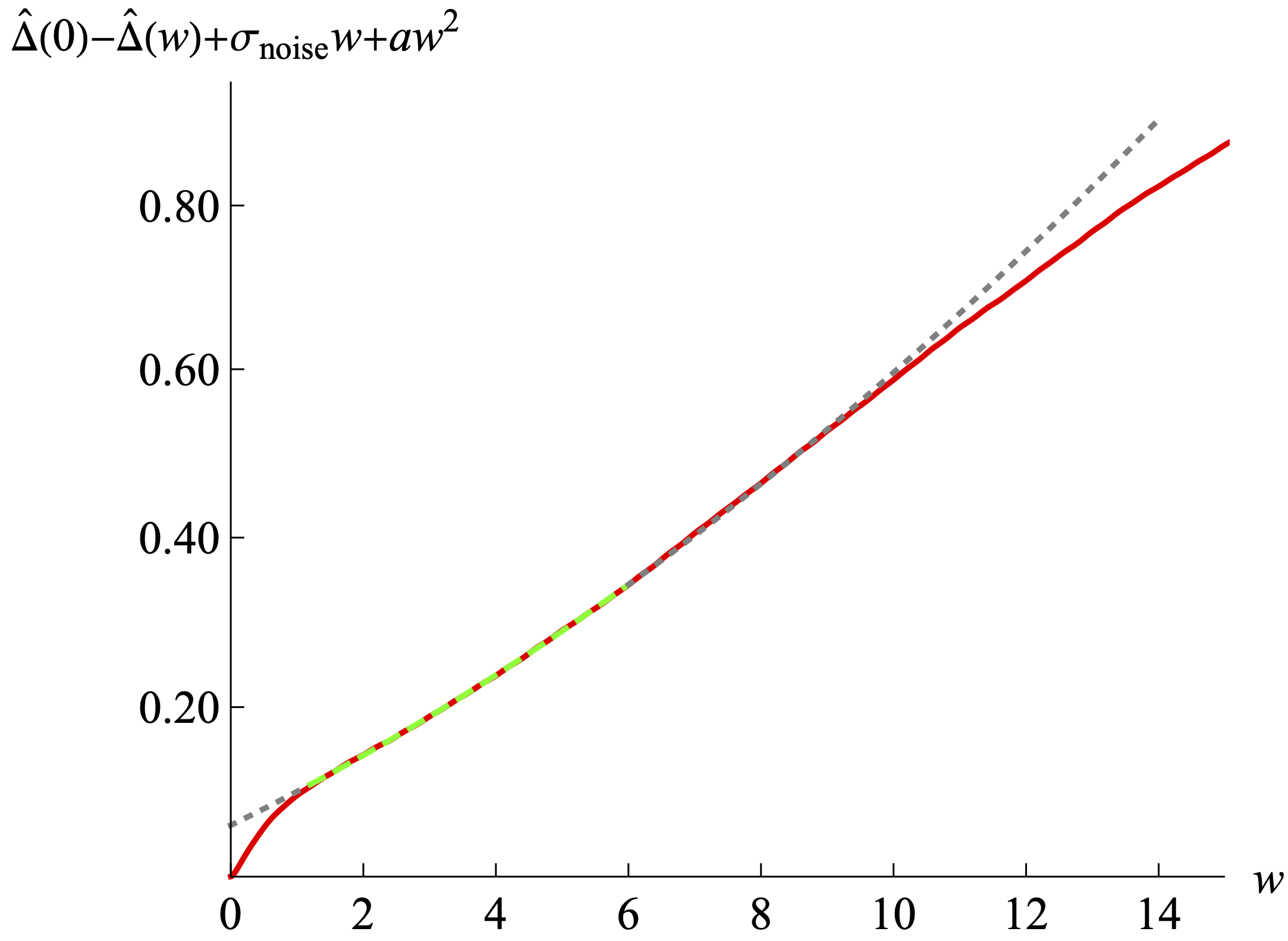}
\caption{Subtraction of the noise contribution $\hat\Delta(0) + \sigma_{\rm noise} |w|+a w^2$ with a small parabolic contribution  (grey dotted, fitting region in green dashed) for    the FeSi ribbon  with LR elasticity and ECs at  $v = 1$. The parabolic term with $a\ge 0$ results from small errors in the procedure of App.~\ref{s:Numerical methods} to estimate the baseline of $\dot u_w$.}
\label{Fig10}
\end{figure}%
\begin{figure*}[b]
\centerline{(a)\hspace{.24\textwidth}(b)\hspace{.24\textwidth}(c)\hspace{.24\textwidth}(d)}
\centerline{     \includegraphics[width=.25\textwidth]{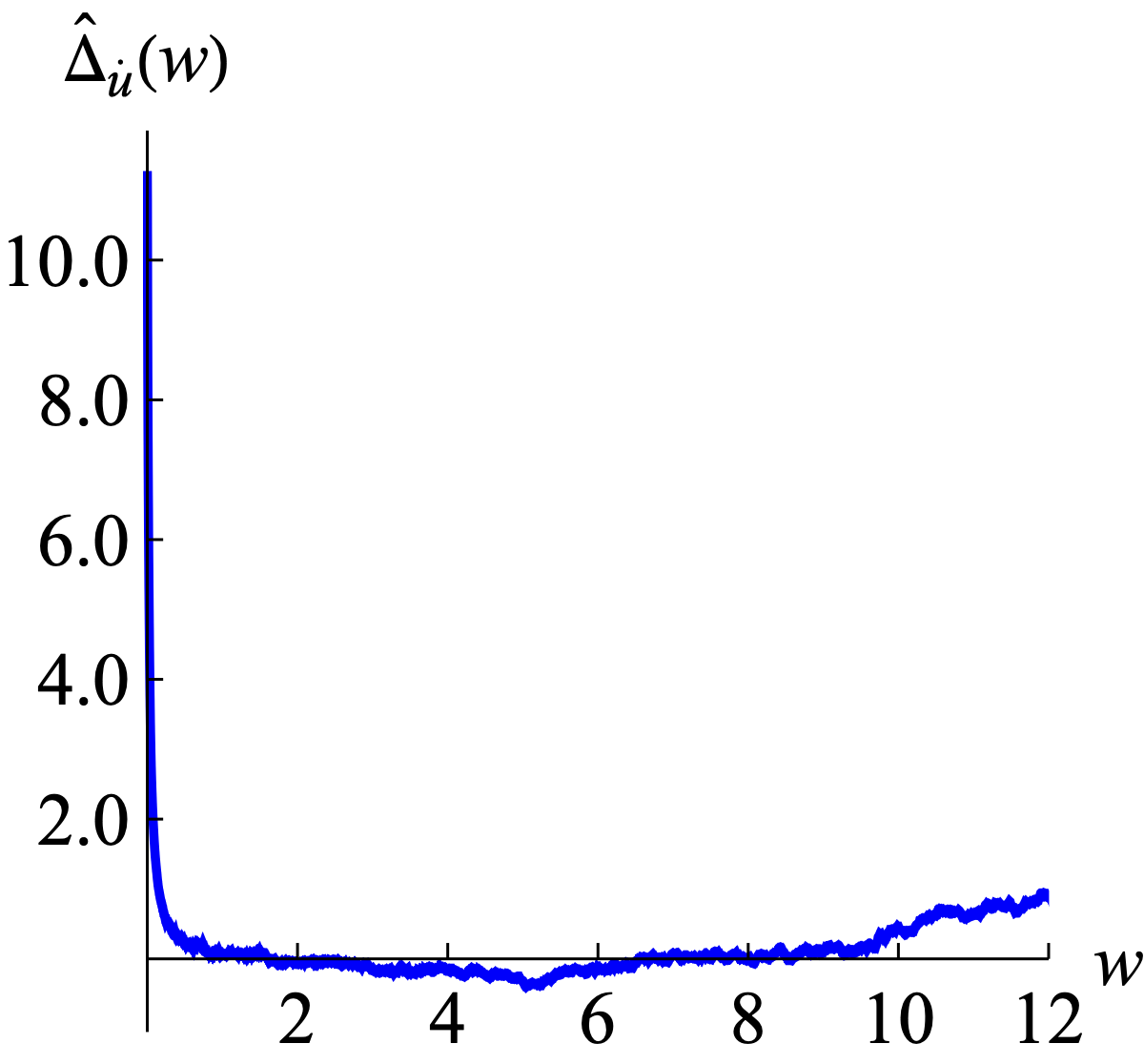}\hfill
    \includegraphics[width=.25\textwidth]{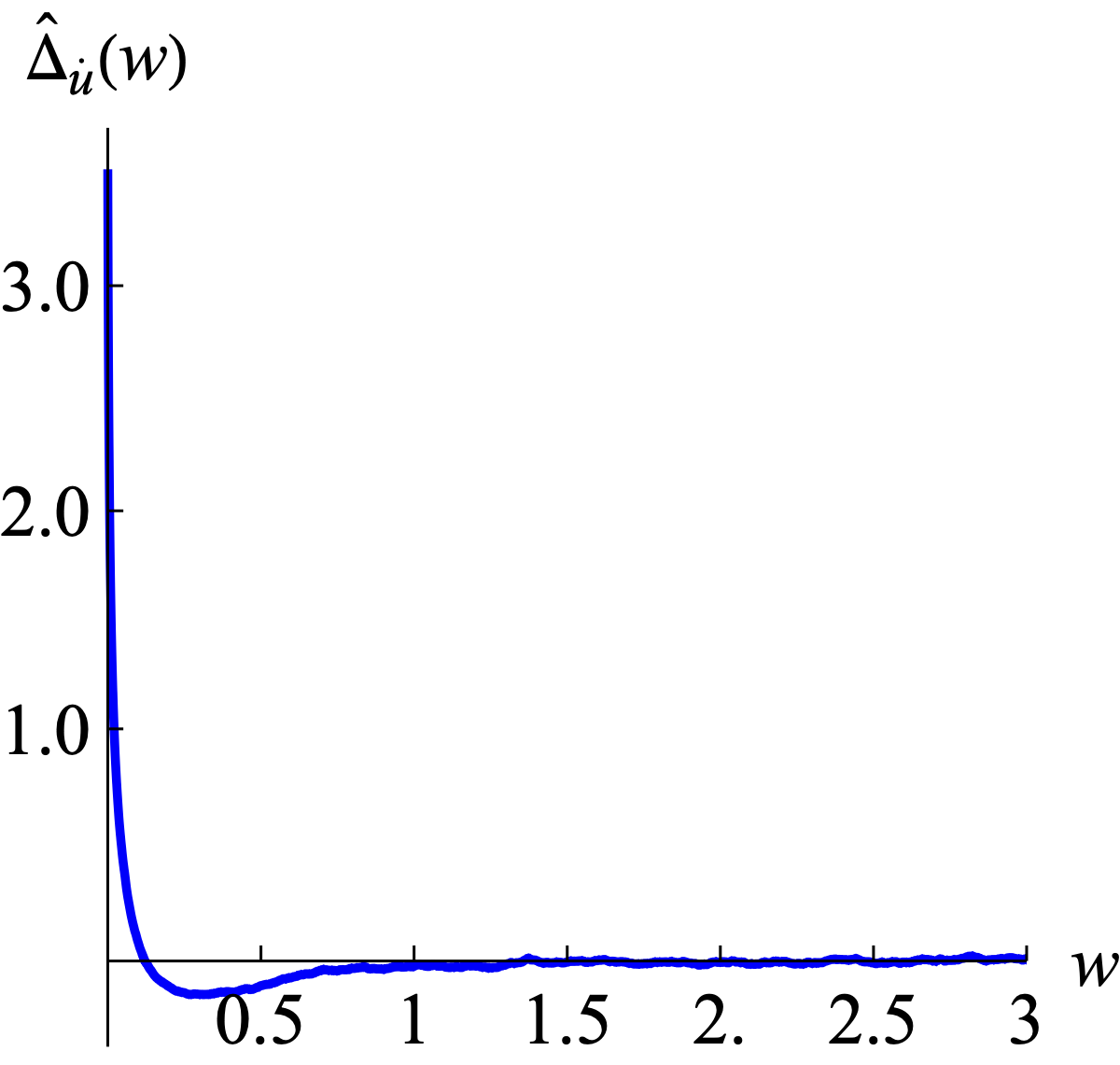}  
 \includegraphics[width=.25\textwidth]{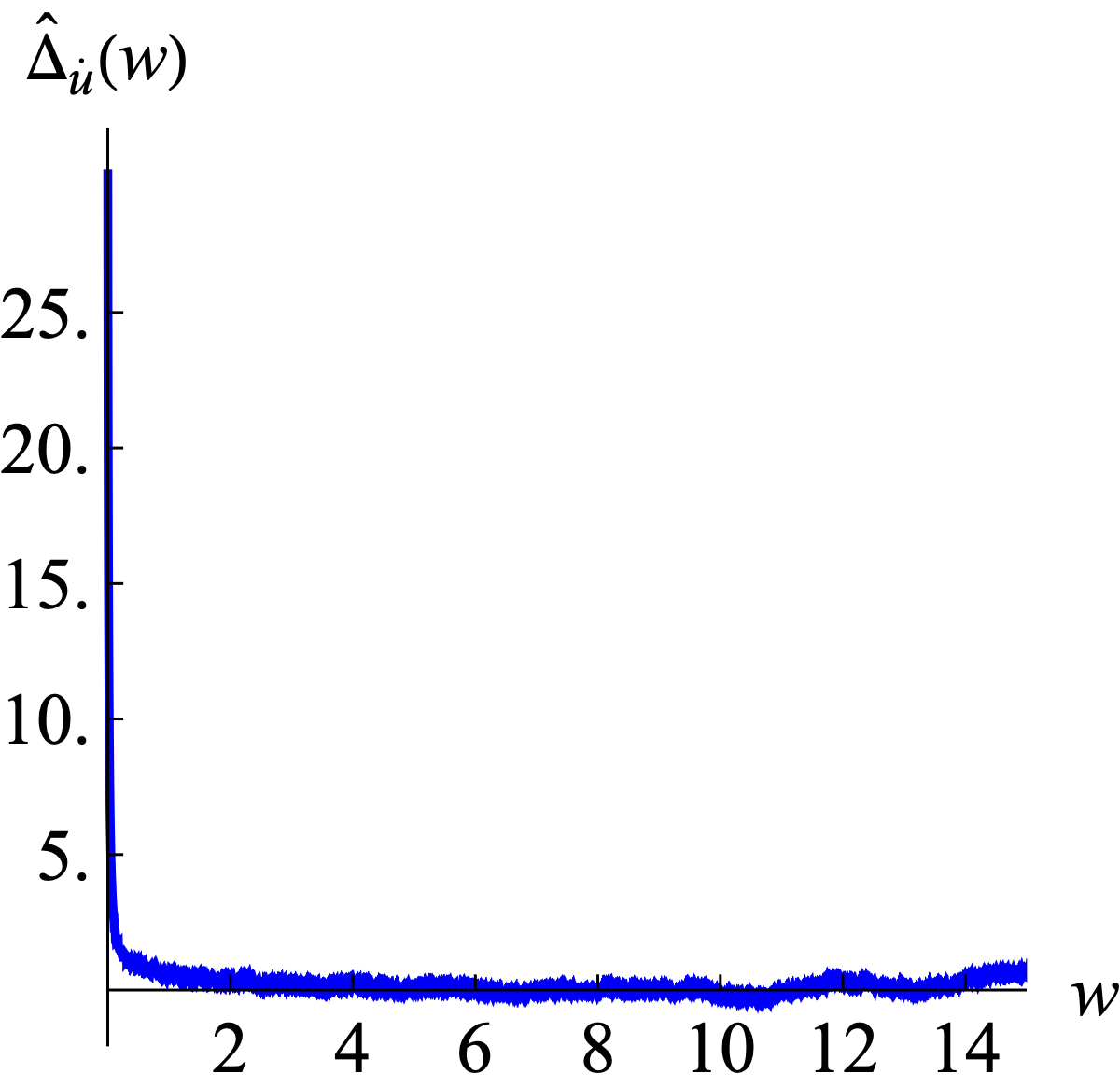}\hfill
    \includegraphics[width=.25\textwidth]{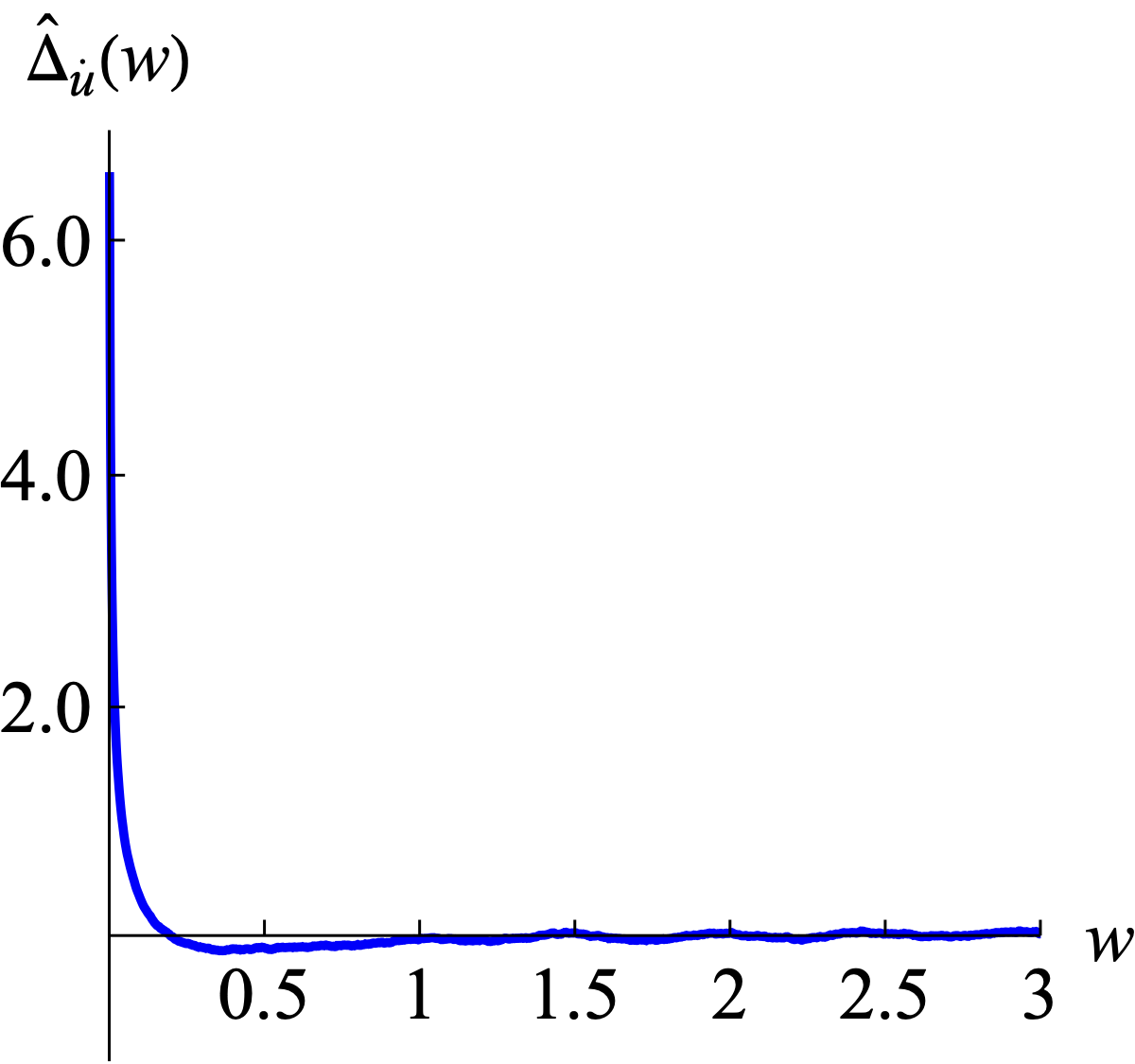}}
\caption{The correlation function $\hat\Delta_{\dot{u}}(w)$ of the domain wall velocity $\dot{u}$.  The scale on which $\hat\Delta_{\dot{u}}(w)$ decays to $0$ is the correlation length $\rho$ of $\hat\Delta(w)$. 
Plots are ordered, as in the main text, for 
({\bf a})    FeSiB film (SR elasticity, no ECs).
 ({\bf b})    FeCoB ribbon (SR elasticity, ECs).  
 ({\bf c})       NiFe film (LR elasticity,   no  ECs).
  ({\bf d})   FeSi ribbon (LR elasticity,     ECs).
 }
\label{Fig8Suppl}
\end{figure*}%
Taking derivatives of the measured function $\hat\Delta_v(w)$ is noisy, but we are in the fortunate position to  have direct access to the velocity correlation function $\hat\Delta_{\dot{u}}(w)$,
\be
\hat\Delta_{\dot{u}}(w-w'): =  \overline{\dot u_w \dot u_{w'}}  = - v^2 \hat\Delta^{\prime \prime }_v(w-w'). 
\ee%
Using this in Eq.~\eqref{OperatorUnfolding1deltau} we get \Eq{OperatorUnfolding1-text} of the main text, \begin{align}
\hat\Delta(w)  =  \hat\Delta_v(w) + \tau^2 \hat\Delta_{\dot{u}}(w) \label{OperatorUnfolding1}.
\end{align}  
In the small-$v$ limit \Eq{DeltaDefRounded-text} can be approximated by a boundary-layer ansatz~\cite{terBurgWiese2020,Wiese2021}, which gives an alternative, robust, albeit less precise,  deconvolution procedure,
\begin{eqnarray}
\hat\Delta_v(w) &=&  \hat\Delta ( \tilde w ) ,\label{boundaryLayer}\\
\tilde{w} &:=& \sqrt{w^2 + (v \tau)^2}.\label{boundaryLayer2}
\end{eqnarray}
Our second strategy to reconstruct $\hat\Delta(w)$ is to plot $\hat\Delta_v(w)$  {\it vs.~}$\tilde{w}$, and determine   $\tau$ which gives the straightest curve at small $\tilde w$. An example is shown in   Fig.~\ref{Fig2unfapp}(a).

\section{Higher-order   deconvolution}
\label{s:Higher-order deconvolution}
We showed that application of \Eq{OperatorUnfolding1} to the measured $\hat\Delta_v(w)$  removes part of the boundary layer $\delta_{w} = v \tau$, but that it creates a new smaller boundary layer of size $\delta_w^{\prime}$.
 On a phenomenological level, we found that  inclusion of an additional term substantially improves the accuracy, 
\bea
\hat\Delta(w)
&= & \hat\Delta_v(w) + \tau^2 \Big[ 1 +  v \tau^\prime \partial_w+ ...\Big]  \hat\Delta_{\dot{u}}(w) . \label{OperatorUnfolding2}
\eea
Such a   term  may arise for a non-exponentially decaying response function $R(t)$. In principle, the procedure can be improved using a second-order derivative in the square brackets. 
 While   a single derivative of $\hat\Delta_{\dot{u}}(w)$ still gives a signal relatively free of noise, adding a second derivative is not possible for our data. In     Fig.~\ref{Fig2unfapp}(b)  we show for the  FeSiB film  the result of the deconvolution \eqref{OperatorUnfolding1} compared to the improved deconvolution    \eqref{OperatorUnfolding2}.

\section{Details for the four samples}   
\label{s:Analysisdetails}

\subsection{FeSiB film: SR   interactions  without ECs}

For the   amorphous FeSiB film with thickness of $200\,$nm,  we show in Fig.~\ref{Fig7Suppl}    the means $\mathcal{M}_i$  for  a single run (in color), compared to the mean $N^{-1}\sum_{i=1}^N \ca M_i$ over all $N$ runs (in black). $\hat\Delta_v(0)$ is extracted from the plateau value at large $w$. Subtracting $\hat\Delta_v(0)$    gives the curve reported in the main text in Fig.~\ref{Fig2}. In Fig.\ \ref{Fig2unfapp} we show for the same sample comparison of   deconvolution  via \Eq{OperatorUnfolding1-text} discussed in the main text,  deconvolution via the boundary layer \eqref{boundaryLayer}-\eqref{boundaryLayer2} and   secondary  deconvolution via \Eq{OperatorUnfolding2}. All procedures are in quantitative  agreement. In our chosen units,  $w=1$  corresponds to $ 2.5\,$ms, assuming a single wall to estimate the driving velocity. Due to the high level of correlation between the walls, we believe this estimation is justified. The number of domain walls is estimated to be  around 3000~\cite{BohnDurinCorreaMachadoDellaPaceChesmanSommer2018}. This implies that  $w=1$  corresponds to   approximately $   1.5\,\rm mm$.

\subsection{FeCoB ribbon: SR   interactions  with ECs}
\label{s:SRWoutECdetails}
For the   amorphous FeCoB ribbon,   Fig.~\ref{Fig9}  shows the subtraction of $\hat\Delta(0)$ plus an additional linear contribution due to white noise as given  in \Eq{DeltaNoisecontr}. All   data presented in the main text are after this subtraction.  In our chosen units,  $w=1$  corresponds to $ 0.2\,$s $\approx 135\,\mu$m. The number of domain walls is estimated to be around 5.

\subsection{NiFe film: LR   interactions  without ECs}   

 deconvolution as presented   in Fig.~\ref{Fig3}(a)  for the  polycrystalline NiFe film having thickness of $200$~nm is done using $\tau= 0.39$.  In our chosen units,  $w=1$  corresponds to $ 2.5\,$ms $ \approx 100\,\mu$m. The number of domain walls is estimated to be around 5000~\cite{SilvaCorreaPaceCidKernCararaChesmanSantosRodriguez-SuarezAzevedoRezendeBohn2017}. 

\subsection{FeSi ribbon: LR   interactions  with ECs}
\label{s:LRWECdetails}
Fig.~\ref{Fig10} shows the subtraction of a linear term plus a   small parabolic contribution for the  polycrystalline FeSi ribbon.  The latter  parabolic contribution arises if our estimate for the baseline of  $ \dot{u} $ for run $i$ still contains a small error, see the discussion after \Eq{FwIntegrated}. 
The  deconvolution shown in the main text in   Fig.~\ref{Fig3}(a)  has been done  using $\tau = 0.055$.
 In our chosen units,  $w=1$  corresponds to $ 50\,$ms $\approx 1.385\,\mu$m. The number of domain walls is estimated to be around 5.

\section{Velocity correlations $\hat\Delta_{\dot u}(w)$}\label{s:Velocity correlations}
Measurements of the velocity correlations $\hat\Delta_{\dot u}(w)$ for   our samples  are shown in Fig.~\ref{Fig8Suppl}. 
The scale on which $\hat\Delta_{\dot u}(w)$ 
decays to zero is the same as the  correlation length $\rho$ of $\hat\Delta(w)$ defined in \Eq{RhoDef}.

\renewcommand{\baselinestretch}{0.1}\normalsize
\parskip=0em
\renewcommand{\baselinestretch}{1.0}\normalsize

\vfill

\end{document}